\documentclass{aa}

\usepackage[version=4]{mhchem}
\usepackage[varg]{txfonts}
\usepackage{graphicx}
\usepackage{siunitx}
\usepackage{amssymb}

\usepackage[colorlinks,bookmarks]{hyperref}
\usepackage{color}
\definecolor{linkblue}{rgb}{0,0,0.8}
\definecolor{linkgreen}{rgb}{0,0.5,0}
\hypersetup{pdfpagemode=UseNone, pdfstartview=FitH, linkcolor=linkblue,citecolor=linkblue, urlcolor=linkblue}

\newcommand{\be}{\begin{equation}}
 \newcommand{\ee}{\end{equation}}
\def \dd{\mathrm{d}}

\usepackage{booktabs}

\def\heii   {{\rm{He \, \textsc{ii}}}}

\def\gsim{\lower.5ex\hbox{\gtsima}} 
\def\lsim{\lower.5ex\hbox{\ltsima}} \def\gtsima{$\; \buildrel > \over 
\sim \;$} \def\ltsima{$\; \buildrel < \over \sim \;$}  \def\gsim{\lower.5ex\hbox{\gtsima}} 
\def\lsim{\lower.5ex\hbox{\ltsima}} 
\newcommand{\angstrom}{\mbox{\normalfont\AA}}
\def\Msun {\rm{M}_{\rm{\odot}}}
\def\Zsun {\rm{Z}_{\rm{\odot}}}

\def\Mstar {M_{*}}

\def \MUV {\rm{M}_{\rm{UV}}}
\def \rhouv {\rho_{\rm{UV}}}

\def \sixteenth{$16^{\rm{th}} \,$}
\def \eightyfourth{$84^{\rm{th}} \,$}
\def\fesc {f_{\rm{esc}}}
\def\mh {M_{\rm{h}}}

\def \fej {f^{\rm{ej}}_*}
\def \feff {f^{\rm{eff}}_*}
\def \fcold {f_{\rm{cold}}}
\def \xion {\xi_{\rm{ion}}}

\def \nionsf {\rm \dot N_{ion}^{sf}}

\usepackage{natbib}
\bibpunct{(}{)}{;}{a}{}{,} 

\begin{document}

\title{Synergising semi-analytical models and hydrodynamical simulations to interpret JWST data from the first billion years} 

\author{Valentin Mauerhofer
  \inst{1}\fnmsep\thanks{\email{v.mauerhofer@rug.nl}}
  \and Pratika Dayal \inst{1}
  \and Martin G. Haehnelt \inst{2}
  \and Taysun Kimm \inst{3}
  \and Joakim Rosdahl \inst{4}
  \and Romain Teyssier \inst{5}}

 \institute{Kapteyn Astronomical Institute, University of Groningen, PO Box 800, 9700 AV Groningen, The Netherlands
 \and Kavli Institute for Cosmology and Institute of Astronomy, Madingley Road, CB3 0HA Cambridge, UK
 \and Department of Astronomy, Yonsei University, 50 Yonsei-ro, Seodaemun-gu, Seoul 03722, Republic of Korea
 \and Universite Claude Bernard Lyon 1, CRAL UMR5574, ENS de Lyon, CNRS, Villeurbanne, F-69622, France
 \and Department of Astrophysical Sciences, Princeton University, Peyton Hall, Princeton, NJ 08544, USA}


\abstract
{The field of high redshift galaxy formation has been revolutionised by the James Webb Space Telescope (JWST), which is yielding unprecedented insights into galaxy assembly at early times. In addition to global statistics, including the redshift evolution of the ultraviolet luminosity function (UV LF) and stellar mass function (SMF), new datasets are providing information on galaxy properties, including the mass-metallicity relation, UV spectral slopes ($\beta$), and ionising photon production efficiencies.}
{In this work our key aim is to understand the physical mechanisms that can simultaneously explain the unexpected abundance of bright galaxies at $z \gsim 11$ as well as the metal enrichment and observed spectral properties of galaxies in the early Universe. We also aim to determine the key sources of reionisation in light of recent data. }
 {We incorporated interstellar medium physics - namely, cold gas fractions and star formation efficiencies - derived from the {\sc sphinx}$^{20}$ high-resolution radiation-hydrodynamics simulation into {\sc delphi}, a semi-analytic model that tracks the assembly of dark matter halos and their baryonic components from $z \sim 4.5-40$. In addition, we explored two methodologies to boost galaxy luminosities at $z \gsim 11$: a stellar initial mass function (IMF) that becomes increasingly top-heavy with decreasing metallicity and increasing redshift (eIMF model), and star formation efficiencies that increase with increasing redshift (eSFE model).}
 {Our key findings are the following: {\it (i)} Both the eIMF and eSFE models can explain the abundance of bright galaxies at $z \gsim 11$. {\it (ii)} Dust attenuation plays an important role in determining the bright end ($\MUV \lsim -21$) of the UV LF at $z \lsim 11$. At higher redshifts, dust is too dispersed to have an impact on the UV luminosity. {\it (iii)} The mass-metallicity relation is in place as early as $z \sim 17$ in all models, although its slope is model-dependent. {\it (iv)} Within the spread of both the models and observations, all of our models are in good agreement with current estimates of $\beta$ slopes at $z \sim 5-17$ and Balmer break strengths at $z \sim 6-10$. {\it (v)} In the eIMF model, galaxies at $z\gsim12$ or with $\MUV\gsim-18$ show typical values of $\xion \sim 10^{25.55}~{\rm [Hz~erg^{-1}]}$, a factor two larger than in other models. {\it (vi)} Star formation in low-mass galaxies ($M_* \lsim 10^{9}\Msun$) is the key reionisation driver, providing the bulk ($\sim 85\%$) of ionising photons down to its midpoint at $z \sim 7$.}
   {}

\titlerunning{Synergising simulations and SAMs to shed light on the first billion years }
\authorrunning{V.Mauerhofer et al.}
\maketitle

\section{Introduction} \label{sec:intro}

The emergence of galaxies in the first billion years of the Universe remains a key frontier in the field of galaxy formation. Representing the seeds of all subsequent structure formation, these early sources were responsible for producing and releasing the first heavy elements and dust grains into the interstellar medium \citep[ISM;][]{Maiolino19}. Crucially, their stars and black holes produced the first hydrogen ionising photons, starting the Epoch of Reionisation (EoR), that marks the last major phase transition of all of the hydrogen in the Universe \citep[for reviews, see e.g.][]{dayal2018,Gnedin22}. However, the emergence of these sources and their large-scale impacts remain key outstanding issues in the field of physical cosmology.

Our view of the first galaxies has evolved dramatically since the launch of the James Webb Space Telescope \citep[JWST; see reviews by][]{Robertson22,Adamo2024}. Its incredible sensitivity has yielded unprecedented insights into galaxy properties well within the EoR, including the redshift evolution of the ultraviolet luminosity function (UV LF) at $z \sim 7-20$ \citep[e.g.][]{ Bouwens23jwst,Naidu22, Adams24, Donnan24, Harikane22a, Harikane23, Harikane24, Leung23, McLeod24, Perez-Gonzales23, Willott24, Kokorev24, Whitler25}, the stellar mass function (SMF) at $z \sim 5-12$ \citep[e.g.][]{Navarro-Carrera24, Weibel24, Harvey25}
ISM metallicities at $z \sim 5-7$ through emission lines \citep[e.g.][]{Curti24a, Chemerynska24}, UV spectral slopes at $z \sim 7-12$ \citep[e.g.][]{Austin24, Franco24, Kokorev24, Rinaldi24}, Balmer break strengths at $z \sim 6-10$ \citep{Vikaeus24} and even hints on hydrogen ionising photon production efficiencies at $z \sim 5-9$ \citep[e.g.][]{Simmonds24,Llerena24, Begley24}. Crucially, it has also revealed an overabundance of bright sources through the UV LF at $z \gtrsim 11$ \citep[e.g.][]{Casey24,Donnan24, Naidu22, Finkelstein24, Robertson24}. Although a few photometrically selected high redshift galaxies were later found to be low-redshift interlopers with strong nebular emission lines \citep{Naidu22schrodinger, Arrabal23}, numerous spectroscopic confirmations by NIRspec have solidified this overabundance of bright sources persisting out to $z \sim 14$ \citep{Carniani24, Harikane24_spec}.

The prevalence of such bright sources within the first billion years has led to a number of theoretical explanations, including {\it (i)} a decrease in dust attenuation with increasing redshift due to dust ejection in radiation-driven outflows \citep{Ferrara23, Fiore23, Ziparo23,nakazato2024}, with dust being more dispersed in the host halo \citep{nikopoulos24} or being spatially segregated from star-forming regions \citep{Ziparo23}; {\it (ii)} a bias in flux-limited observations that leads to the preferential detection of galaxies with extreme star-bursts or stochastic star formation \citep{Mason23, Mirocha23, Shen23, Sun23,nikopoulos24}; {\it (iii)} accreting black holes contributing to the UV luminosity \citep{Pacucci22, dayal2024, Hegde24}; {\it (iv)} largely inefficient stellar feedback, which could boost the star formation efficiency \citep{Dekel23, Li24}; and {\it (v)} a stellar initial mass function (IMF) that becomes increasingly top-heavy with increasing redshift, resulting in high light-to-mass ratios \citep{Pacucci22, Haslbauer2022, Trinca24, Cueto24, Hutter24, Yung2024,jeong2024,Lu25}. This last option is extremely promising, given a number of recent observations that seem to hint at a top-heavy IMF at early epochs through enhanced ionising photon production rates \citep{Simmonds24}, extremely blue UV slopes \citep{Topping22, Cullen24, Morales24}, and elevated [N/O] ratios \citep{Bunker23, Cameron23, Isobe23, Curti24b, Watanabe24}. Additionally, numerical simulations following the birth of star clusters at high redshifts show that elevated gas densities and low metallicities - conditions typical of the systems detected by JWST - favour top-heavy IMFs \citep{Chon22, Chon24}. Finally,  the high values observed for the [OIII]$_{\rm 88\mu m}$ to [CII]$_{\rm 158\mu m}$ ratio at $z>6$ \citep{Carniani20, Harikane20} have been used to infer a low metallicity stellar population with a top-heavy IMF \citep{Katz22}.

In this work, we explore several of these mechanisms within the semi-analytical {\sc delphi} framework aimed at tracking the assembly of dark matter halos and their baryons in the first billion years \citep{dayal2014, dayal2022, Mauerhofer23}. Compared to the latest version of {\sc delphi} \citep{Mauerhofer23}, referred to as {\sc delphi}23 henceforth, we include the following key enhancements in this work. Firstly, we introduce a stochastic component into the star formation modelling by extracting ISM properties of galaxies from {\sc sphinx}$^{20}$, a state-of-the-art radiation hydrodynamics (RHD) simulation \citep{Rosdahl22}. The high-resolution, accurate out-of-equilibrium ISM photochemistry and the turbulence-based star formation recipe of this large cosmological simulation provide us with distributions of star formation efficiencies and cold gas fractions as a function of redshift and halo mass. Stochasticity is automatically included in these distributions, since the simulated galaxies' star formation properties are naturally bursty. Sampling {\sc sphinx}$^{20}$ distributions allows us to introduce random scatter in {\sc delphi}, with some rare galaxies experiencing starbursts. Secondly, we implement a version of the model with an evolving IMF that becomes increasingly top-heavy with decreasing metallicities and increasing redshifts \citep[e.g.][]{Cueto24}. With these new prescriptions in hand, we aim to analyse the conditions required to reproduce the overabundance of bright galaxies at $z>9$, in addition to studying the physical and observable properties of these galaxy populations. 

Throughout this work, we assume a $\Lambda$CDM model with dark energy, dark matter, and baryonic densities in units of the critical density as $\Omega_{\Lambda}= 0.6889$, $\Omega_{m}= 0.3097$, and $\Omega_{b}= 0.04897$, respectively. The Hubble constant we adopt is $H_0=100\, h\,{\rm km}\,{\rm s}^{-1}\,{\rm Mpc}^{-1}$, with $h=0.677$, a spectral index of $n=0.965$, and a normalisation of $\sigma_{8}=0.811$ \citep[][]{planck2018}. All units in this work are comoving units, and magnitudes are in the standard AB system \citep{ABmag}.

The paper is organised as follows. In Sect. \ref{sec:method}, we describe the methodology of {\sc delphi}, including the dark matter merger trees, the derivation of ISM properties from {\sc sphinx}$^{20}$, the modelling of star formation and its associated feedback, the metal and dust enrichment of early systems, and the emerging luminosity for models of varying IMF and star formation efficiency. We undertake a thorough comparison of the properties of these systems against observables - including the redshift evolution of the UV LF and the SMF, dust and metal masses, UV spectral slopes, Balmer break strengths, and the ionising photon production efficiency - in Sect. \ref{sec:predictions}. We then analyse the reionisation history and its key sources for a number of escape fraction scenarios in Sect. \ref{sec:reionisation}. We end by summarising our key findings, and discussing caveats and possible improvements in Sect. \ref{sec:summary}.

\section{The theoretical model} \label{sec:method}

The new approach we adopt here combines the {\sc delphi} semi-analytic galaxy formation model with insights on the ISM - in terms of the cold gas fraction and stochastic star formation - from the (very high-resolution) {\sc sphinx}$^{20}$ radiation hydrodynamics simulation \citep{Rosdahl22}. The models are briefly described here, and readers are referred to previous works describing {\sc delphi} \citep{dayal2014,dayal2022, Mauerhofer23} and {\sc sphinx}$^{20}$ \citep{Ramses,Joki2013,RamsesRT} for complete details. Finally, we use a Kroupa IMF \citep{Kroupa01} throughout this work, with a slope of -1.3 (-2.3) between 0.1-0.5 (0.5-100) $\Msun$, except for in the evolving IMF model (Sect. \ref{sec:evol_imf}).

\subsection{Merger trees and gas assembly} \label{sec:merger_tree}
We start by building dark matter merger trees, that form the basis of {\sc delphi}, for 600 halos at $z=4.5$ with (logarithmically equally spaced) masses between $\mh = 10^{8-14} \, \Msun$ using a binary merger tree algorithm \citep{parkinson2008}. In brief, the merger tree for each simulated halo starts at a redshift of $z=4.5$ and runs backwards in time up to $z=40.8$, with each halo fragmenting into its progenitors. A halo of mass $M_0$ can either fragment into two halos with masses larger than the resolution mass $M_{\rm{res}} < M < M_0/2$, or lose a fraction of its mass, meaning fragment into progenitor halo(s) below the mass resolution. In the latter case, the mass brought in by unresolved progenitors is denoted as 'smooth accretion' from the intergalactic medium (IGM). The merger tree used in this work stores outputs over 44 time-steps equally spaced by 30 Myr with a resolution mass of $M_{\rm{res}} =10^{8.0} \, \Msun$. Further, each halo at $z =4.5$ is assigned a number density by matching to the Sheth-Tormen \citep{sheth1999} halo mass function (HMF). All the progenitors of a halo are assigned the same number density as their $z=4.5$ descendant; we have confirmed the resulting HMFs are in accordance with the Sheth-Tormen HMFs at all higher redshifts.

\begin{figure*} 
  \resizebox{\hsize}{!}{\includegraphics{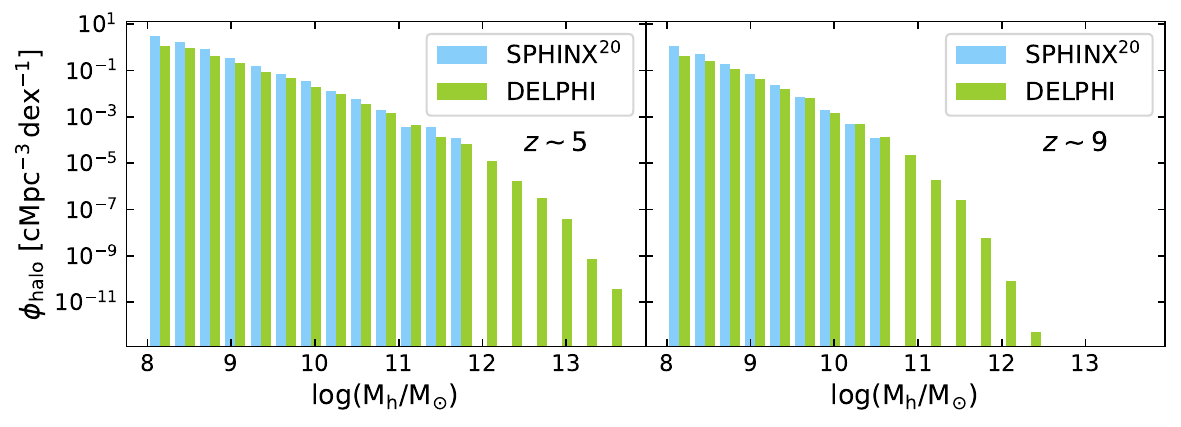}}
  \caption{Halo mass functions obtained from the {\sc sphinx}$^{20}$ and {\sc delphi} models, shown using blue and green histograms, respectively, as marked. The {\it left} and {\it right} panels show results at $z\sim 5$ and $9$, respectively.}
  \label{fig:hist_halo_mass}
\end{figure*}

In terms of baryons, the first halos ('starting leaves') of any merger tree are assigned an initial gas mass that is linked to the halo mass through the cosmological baryon-to-dark matter ratio such that $M_{\rm{g}}^{\rm i} = (\Omega_b/\Omega_m) M_h$. If any halo has progenitors, its initial gas mass is the sum of the final gas mass brought in by merging progenitors - after star formation and the resulting supernova (SN) feedback (Sect. \ref{sec:sf-feedback}) - and the gas mass smoothly accreted from the IGM. For the latter, we reasonably assume smoothly accreted dark matter to drag in a cosmological ratio of baryons. 

\subsection{Cold gas fractions and star formation efficiencies from the {\sc sphinx}$^{20}$ simulation} \label{sec:sphinx}
In this work we incorporate the cold gas fraction and star formation efficiency (SFE) distributions sampled from the {\sc sphinx}$^{20}$ simulation\footnote{\url{https://sphinx.univ-lyon1.fr/}} into the {\sc delphi} semi-analytic model. In brief, {\sc sphinx}$^{20}$ simulates a cubic volume of (20 cMpc)$^3$ using the \textsc{Ramses-RT} adaptive mesh refinement (AMR) code. With a dark matter particle resolution mass of $2.5 \times 10^5 \Msun$, this simulation resolves halos at the atomic cooling mass of $3 \times 10^7\Msun$ with 120 dark matter particles. The volume simulated here was selected as the most average amongst 60 cosmological initial conditions in order to minimise cosmic variance biases. The resolution reaches 10 pc in the densest regions of the ISM, sufficient to resolve gas fragmentation into multiple star-forming regions, the escape of ionising radiation and the photochemistry of interstellar gas.

In terms of gravity and hydrodynamics, the collisionless dark matter particles and stellar particles are evolved using a particle-mesh solver with cloud-in-cell interpolation \citep{Guillet11} and the gas evolution is computed by solving the Euler equations with a second-order Godunov scheme using the HLLC (Harten–Lax–van Leer contact) Riemann solver \citep{Toro94}. Gas is converted into stars in cells that have a gas density higher than 10 $\rm cm^{-3}$, a locally turbulent Jeans length that is smaller than the cell width and gas that is locally convergent \citep{Rosdahl18}. When a cell satisfies those conditions, its star formation efficiency per free-fall time is computed as a function of the local virial parameter and turbulence \citep{Federrath12}, and its gas converted stochastically into stellar particles \citep{Rasera06} with an initial mass of 400$\Msun$. The feedback implementation of {\sc sphinx}$^{20}$, which is crucial for shaping the cold gas fraction and star formation histories, follows the prescription detailed in \cite{Rosdahl18} where each SN explosion injects $10^{51}$ erg of energy 'mechanically' into the ISM \citep{Kimm15}. Individual stellar particles undergo several SN explosions from ages 3 to 50 Myr, with an average of four SN explosions per 100$\Msun$ formed. This SN rate is artificially boosted by a factor of roughly four compared to a Kroupa IMF, which is necessary to suppress star formation and produce realistic high redshift luminosity functions \citep{Rosdahl22}. The code includes radiative transfer (RT) in two bins of hydrogen-ionising photons using a first-order moment method. RT is an essential ingredient to obtain an accurate non-equilibrium ionisation state and the temperature of hydrogen, crucial for the cold gas fraction considered in this work. The ionising photons are emitted from stellar particles within the simulated galaxies using the BPASSv2.2.1 \citep{BPASS1,BPASS2} spectral energy distribution (SED) model. 

\begin{figure*} 
  \resizebox{\hsize}{!}{\includegraphics{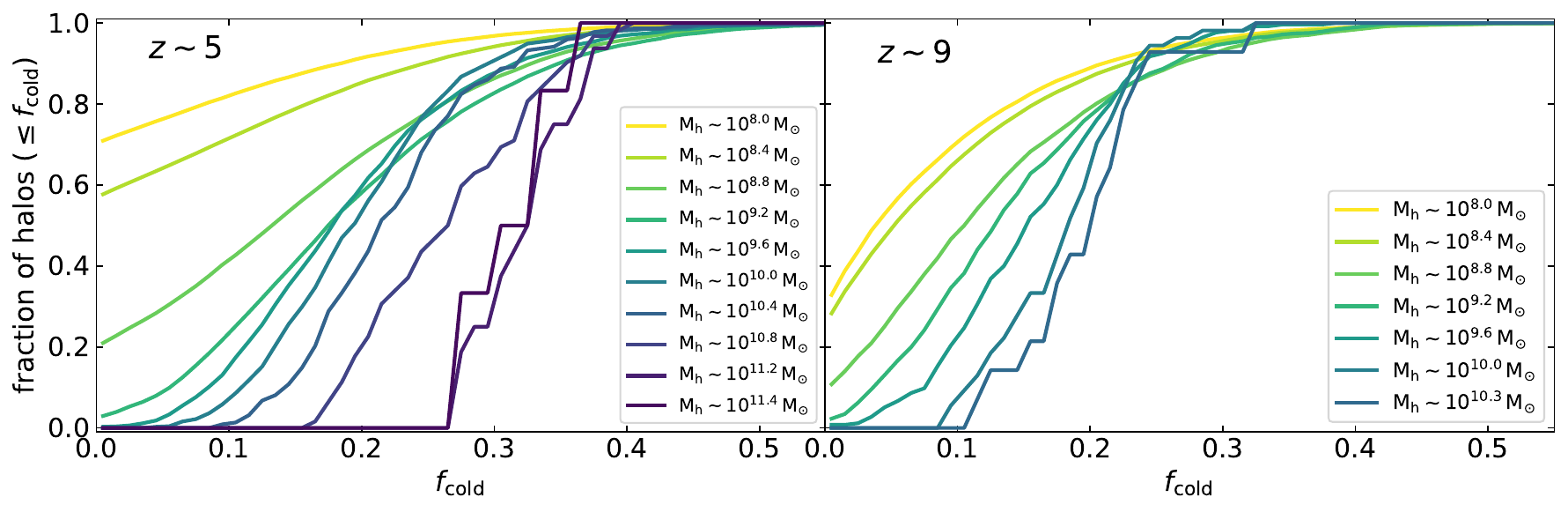}}
  \caption{Fraction of halos with a cold gas fraction ($\fcold$) below the value shown on the horizontal axis from the {\sc sphinx}$^{20}$ simulations, for the mass bins marked, at $z=5$ ({\it left panel}) and $z=9$ ({\it right panel}). The cold gas fraction is defined as the fraction of gas mass that is below 1000 K, within the virial radius of the halo.}
  \label{fig:cold_gas_distribution}
\end{figure*}

\begin{figure*} 
  \resizebox{\hsize}{!}{\includegraphics{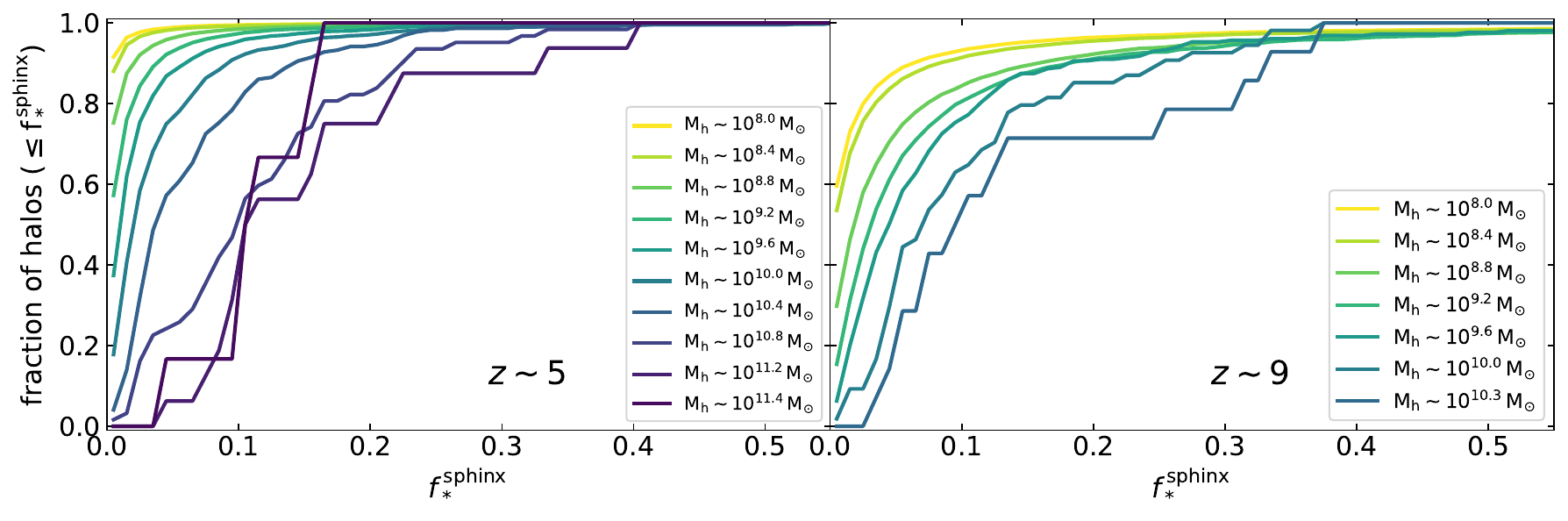}}
  \caption{Fraction of halos with a star formation efficiency value ($f_*^{\rm{sphinx}}$) below the value shown on the horizontal axis from the {\sc sphinx}$^{20}$ simulations, for the mass bins marked, at $z=5$ ({\it left panel}) and $z=9$ ({\it right panel}). The star formation efficiency is defined as the ratio between the mass of stars younger than 30 Myr and the cold gas mass; if a halo has no cold gas mass, we define its star formation efficiency to be $f_*^{\rm{sphinx}}=0$.} 
  \label{fig:f_star_distribution}
\end{figure*}

The most massive halo in {\sc sphinx}$^{20}$ has a virial mass of $4.3 \times 10^{11} \Msun$, $9.5 \times 10^{10} \Msun$, $2.8 \times 10^{10} \Msun$ and $4.0 \times 10^{9} \Msun$ at $z \sim 5, 7, 9 $ and $12$, respectively. In this paper, we use all the simulated halos from {\sc sphinx}$^{20}$ more massive than $10^8 \,\Msun$, corresponding to the lowest halo mass in the merger tree for {\sc delphi}, in eleven snapshots from redshift 5 to 15. This results in a total of \num{48801}, \num{31338}, \num{15463} and \num{3481} {\sc sphinx}$^{20}$ sources at $z \sim 5, 7, 9$ and 12, respectively. 
We note that at any redshift, as expected from hierarchical structure formation, the HMFs from both the {\sc delphi} merger trees and {\sc sphinx}$^{20}$ follow the same shapes, as shown at $z \sim 5$ and $9$ in Fig. \ref{fig:hist_halo_mass}. However, given its analytic nature, at any redshift, the {\sc delphi} merger trees contain halos that are up to two orders of magnitude more massive than the most massive {\sc sphinx}$^{20}$ halos.

In this new version of {\sc delphi}, the first major change is that we split the total gas mass into warm and cold components. We define cold gas as that with a temperature $<1000$K - this forms both the reservoir for star formation and grain growth of dust grains in the ISM (Sect. \ref{sec:dust}). The temperature threshold of 1000K is chosen to demarcate gas that is cold enough to form stars in some {\sc sphinx}$^{20}$ cells, but not cold enough to enter the regime of very dense molecular clouds that cannot be resolved by the simulation. We then assign a cold gas mass fraction, $\fcold$, to each halo in {\sc delphi} by sampling halo properties from the {\sc sphinx}$^{20}$ simulation, which introduces a stochastic component in our model. To do so, we sample the probability distribution function (PDFs) for the cold gas fraction in {\sc sphinx}$^{20}$ in stellar mass bins separated by 0.1 dex ranging from $10^{8} \Msun$ up to the largest mass at that redshift. The bins partially overlap so as to avoid discontinuities between neighbouring distributions. The PDFs of this cold gas fraction are plotted in Fig. \ref{fig:cold_gas_distribution}, for a subset of halo mass bins and two redshifts, $z \sim 5$ and $9$. As seen from this figure, at $z\sim 5$, about $70\%~(58\%)$ of the low-mass halos with $\mh \sim 10^{8} ~ (10^{8.4})\Msun$ have no cold gas. This is, in part, because they reside in ionised regions where the UV background from reionisation prevents gas cooling and/or accretion of cold gas \citep[see][]{Katz20}, and, in part due to internal feedback processes. Only $10-20\%$ of such low-mass halos have values of $\fcold \sim 0.2-0.4$. An increasing fraction of halos start hosting high cold gas fractions with increasing halo mass at $z \sim 5$. For example, roughly $50\%$ of the halos with $\mh \gsim 10^{10.4}\Msun$ show $\fcold \sim 0.2-0.45$. While roughly the same trends persist at $z \sim 9$, a key difference is that low-mass halos ($\mh \lsim 10^{8.4}\Msun$) contain larger cold gas fractions than at lower redshifts. For example, only 35\% of such halos do not contain any cold gas, with about 65\% showing values of $\fcold \sim 0.05-0.3$. This can be explained by the diminution of the UV background compared to that at $z\sim 5$. The behaviour at larger masses is roughly the same across these two redshifts, with about $50\%$ of $\mh \gsim 10^{10}\Msun$ halos showing values of $\fcold \sim 0.2-0.4$.  

The second key change in our new version of {\sc delphi} is the inclusion of a  stochastic star formation parameter, $f_*^{\rm{sphinx}}$. This is the ratio of the stellar mass formed in the last 30 Myr ($M_*[<30 \, \rm{Myr}]$) divided by the current mass of cold ($<1000$ K) gas; the time-step of 30 Myr is chosen to be coherent with the time-stepping of the underlying {\sc delphi} merger tree. If the cold gas mass of a halo is zero, we define $f_*^{\rm{sphinx}} = 0$, and we cap $f_*^{\rm{sphinx}}$ to a maximum of 1. 
To summarise,
\be 
f_*^{\rm{sphinx}} = \min \left(1, \frac{M_*[<30 \, \rm{Myr}]}{M_{\rm{g}}^{\rm i} \fcold} \right).
\ee
The resulting efficiencies, for a number of stellar mass bins, are displayed in Fig. \ref{fig:f_star_distribution}. At $z \sim 5$, about 90\% of $\mh \lsim 10^{8.4}\Msun$ halos show  $f_*^{\rm{sphinx}}=0$, meaning that either their cold gas fraction is 0 or that they did not form stars in the last 30 Myr. Again, the star formation efficiency increases with an increase in the host halo mass with roughly 50\% of $\mh \gsim 10^{10.4}\Msun$ showing $f_*^{\rm{sphinx}} \sim ~ 0.1-0.4$. At $z\sim 9$, a larger fraction ($\sim 40\%$) of low-mass halos with $\mh \lsim 10^{8.4}\Msun$ show non-zero values of $f_*^{\rm{sphinx}}$, driven by their larger cold gas fractions discussed above. In accordance with $z \sim 5$ results, about 50\% of the most massive halos with $\mh \gsim 10^{10} \Msun$ at this redshift show $f_*^{\rm{sphinx}} \sim 0.08-0.4$. 

To assign $\fcold$ and $f_*^{\rm{sphinx}}$ values to {\sc delphi} halos, we choose the {\sc sphinx} probability distribution at the closest redshift and halo mass bin using a standard Monte Carlo sampling. As shown in Fig. \ref{fig:hist_halo_mass} above, {\sc delphi} halos reach higher masses than those in the {\sc sphinx}$^{20}$ simulations given the latter's relatively small box size; as a result, the most massive {\sc delphi} halos have to be matched with smaller {\sc sphinx}$^{20}$ halos. However, both the $\fcold$ and $f_*^{\rm{sphinx}}$ distributions start showing convergence of the PDF for the few most massive {\sc sphinx}$^{20}$ halos. Thus, we expect that matching {\sc delphi} halos with lower mass systems from {\sc sphinx}$^{20}$ does not induce a large error in our results.

\subsection{Star formation rates and supernova feedback} \label{sec:sf-feedback}

The newly formed stellar mass at any time-step can be written as
\be 
\Delta M_*(z) = \feff \times M_{\rm{g}}^{\rm{i}}, \label{eq:stars_formed}
\ee
where the 'effective' star formation efficiency is $\feff = \min(f_{\rm{cold}} \cdot f_*^{\rm{sphinx}}, \fej)$. Here, $M_{\rm{g}}^{\rm{i}}$ is the initial gas mass in the halo at the start of the time-step. Further, $\fej$ is the star formation efficiency that produces enough Type II Supernova (SNII) feedback to eject the rest of the gas \citep[see][]{Mauerhofer23}:
\be
\fej = \frac{v_c^2}{v_c^2 + f_w \, \nu \, E_{51}}, \label{eq:fej}
\ee
where $v_c$ is the halo rotational velocity, $\nu$ is the supernova rate, $E_{51}=10^{51}$erg is the energy per SNII, and $f_w$ is a free parameter that determines the fraction of SN energy coupling to the gas content. If $\fej < f_{\rm{cold}} \cdot f_*^{\rm{sphinx}}$, a galaxy is 'feedback limited'; that is, it loses all its gas due to SNII feedback. In this paper, we fix $f_w = 4\%$, in order to match the observed redshift evolution of the UV LF at $z\lsim 10$, as also detailed in Table \ref{table:model_summaries}. 

\subsection{Dust and metal enrichment} \label{sec:dust}

The evolution of the metal ($M_Z$) and dust contents ($M_d$) of early galaxies are inextricably coupled through the processes of metal/dust production, astration into further star formation, ejection in outflows, dilution in (metal and dust poor) inflows of gas from the IGM, and dust destruction and growth in the ISM. We assume metals, dust and gas to be perfectly mixed in the ISM. In terms of metal production, we assume all SN and AGB (asymptotic giant branch) stars between $2-50\Msun$ to contribute to the metal yield using the results presented in \citet{Kobayashi20}. The lower limit to this mass is driven by the fact that that stars less than 2 $\Msun$ need $\sim 1.7$ Gyr to end their lives, which is larger than the age of the Universe at $z >4.5$; stars more massive than $50\Msun$ are assumed to collapse to black holes without any yield. We consider SNII as the key dust factories, with each SN producing $0.5 \Msun$ of dust \citep[e.g.][]{lesniewska2019,dayal2022}. The coupled evolution of dust and gas-phase metals during a {\sc delphi} time-step is modelled as \citep{dayal2022}

\be \label{eq:MZ}
\frac{\dd M_{Z}}{\dd t} = \dot{M}_{Z}^{\rm{pro}} - \dot{M}_{Z}^{\rm{eje}} - \dot{M}_{Z}^{\rm{ast}} - \dot{M}_d^{\rm{gro}} + \dot{M}_d^{\rm{des}} \\
\ee
\be \label{eq:Md}
\frac{\dd M_d}{\dd t} = \dot{M}_d^{\rm{pro}}  - \dot{M}_d^{\rm{eje}} - \dot{M}_d^{\rm{ast}} + \dot{M}_d^{\rm{gro}}- \dot{M}_d^{\rm{des}}.
\ee
Eqs. \ref{eq:MZ} and \ref{eq:Md} show the time evolution of the metal and dust masses, respectively, considering the processes of production ({\it pro}), ejection ({\it eje}), astration ({\it ast}), the loss of metals into dust grains growing in the cold ISM ({\it gro}), and the destruction of dust grains in the warm ISM that add to the metal content ({\it des}). We use the newly introduced cold gas fraction to calculate the destruction and growth of dust grains in the warm and cold ISM, respectively; this is a key difference with respect to the {\sc delphi}23 model where we had assumed an equi-partition of the gas into the cold and warm phases. 

\subsection{Ultraviolet and ionising luminosities} \label{sec:def_UV}

In order to calculate the intrinsic SED and the hydrogen ionising photon production rate for each galaxy, we convolve its star formation history (assuming the stellar mass in any 30 Myr time-step to have formed through constant star formation) with the Starburst99 (SB99) stellar population synthesis code \citep{leitherer1999} using the 'Geneva High mass loss' tracks \citep{Meynet94}. 

To include the effect of dust, we start by calculating the dust attenuation of UV photons (1450-$1550\,\angstrom$ in the rest-frame) as in \cite{Mauerhofer23}: we use a single grain size of $a=0.05\, \mu m$ and a material density of $s=2.25\, {\rm g\, cm^{-3}}$ appropriate for graphite/carbonaceous grains \citep{Todini01, Nozawa03}. We then assume this dust to be mixed with gas in a radius given by 
\be
r_{\rm{gas}} = 4.5 \lambda r_{\rm{vir}} \times \left(\frac{1+z}{6}\right), \label{eq:rgas}
\ee 
where $\lambda = 0.04$ is the spin parameter \citep[e.g.][]{bullock2001}. As seen, both gas and dust cover an increasing volume fraction of the halo with increasing redshift, as indicated by recent ALMA observations \citep{Fujimoto20, Fudamoto22}. This is used to obtain the dust optical depth to UV photons as $\tau_c = 3 M_{d}[4 \pi r_{\rm{gas}}^2 a s]^{-1}$ yielding the observed UV magnitude. We then use the Calzetti extinction curve \citep{calzetti2000} convolved with this UV optical depth to account for the dust attenuation of the entire SED. This dust attenuated SED is then used to calculate $\beta$ slopes (Sect. \ref{sec:beta}) and the Balmer break strength (Sect. \ref{sec:bb}) in order to compare our models to observations. 

\subsection{Two extreme star formation models}

Besides the fiducial model described above, we also explore two alternative models aimed at increasing the UV luminosities of galaxies at very high redshifts. This is motivated by the overabundance of bright sources detected by JWST at $z \gsim 11$, as discussed in what follows.
\subsubsection{A redshift- and metallicity-dependent initial mass function} \label{sec:evol_imf}

One way of boosting the UV luminosities of galaxies at $z \gsim 10$ is by appealing to an evolving IMF (the 'eIMF' model) that becomes increasingly top-heavy with redshift or decreasing metallicity as supported by high-resolution star formation simulations \citep[e.g.][]{Chon22}. Our modified evolving IMF has the following form, inspired by previous results \citep{Chon22,Cueto24}:
\begin{eqnarray}
    \xi(M) \propto 
    \begin{cases}
    \left(1 - f_\mathrm{massive}\right)\ M^{-1.3} & ,\ 0.1\Msun \leq M < 0.5\Msun\\
        \left(1 - f_\mathrm{massive}\right)\ M^{-2.3} & ,0.5\Msun \leq M < M_\mathrm{c}\\
        f_\mathrm{massive}\ M^{-1} & ,\ M_\mathrm{c} \leq M \leq 100\Msun.
    \end{cases}
    \label{eq_evolvingIMF}
\end{eqnarray}

\begin{figure} 
  \resizebox{\hsize}{!}{\includegraphics{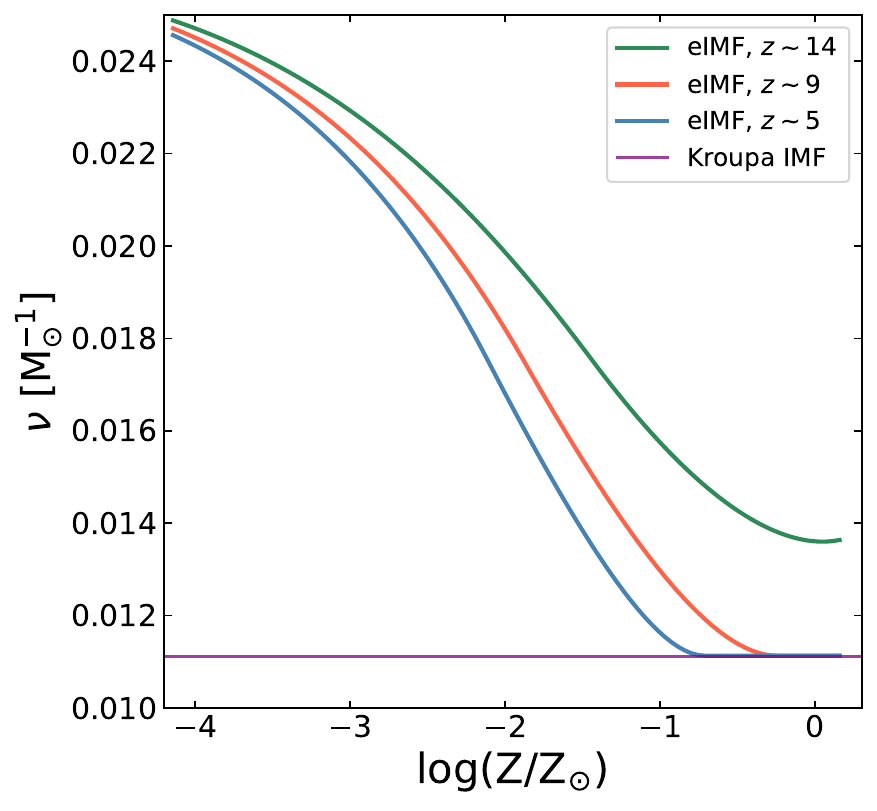}}
  \caption{Evolution of the SN rate ($\nu$) as a function of metallicity, at the redshifts marked, for the eIMF model (Sect. \ref{sec:evol_imf}). The shape of the IMF in this case is determined by the metallicity and redshift. 
 The purple line shows the supernova rate for the fiducial Kroupa IMF.}
  \label{fig:sn_rate}
\end{figure}

Here, $M_\mathrm{c}$ is the cut-off mass at which the Kroupa IMF transitions to a log-flat component and $f_\mathrm{massive}$ is the fraction of stellar mass above $M_\mathrm{c}$. The value of $f_{\rm{massive}}$ is given by \citep{Chon22}:
\be \label{eq:f_massive}
f_{\rm{massive}} = 1.07 \cdot (1-2^x) + 0.04 \cdot 2.67^x \cdot z,
\ee
where $z$ is the redshift and $x=1+\log_{10}(Z/Z_{\odot})$. Here, $Z$ is the metallicity before star formation and $Z_{\odot}=0.0134$ represents the solar metallicity value \citep{Asplund09}. $M_c$ is computed by solving the system of equations formed by the two continuity equations of the IMF at $0.5\Msun$ and at $M_c$, the normalisation equation requiring $\int_{0.1}^{100} \xi(m) \, m \, \dd m = 1$, and the definition of $f_{\rm{massive}} = \int_{M_c}^{100} \xi(m) \, m \, \dd m$. We cap $f_{\rm{massive}}$ below 0.995 to ensure that $M_c$ is always above $0.5 \, \Msun$.

\begin{figure} 
\begin{center}
\center{\includegraphics[scale=0.6]{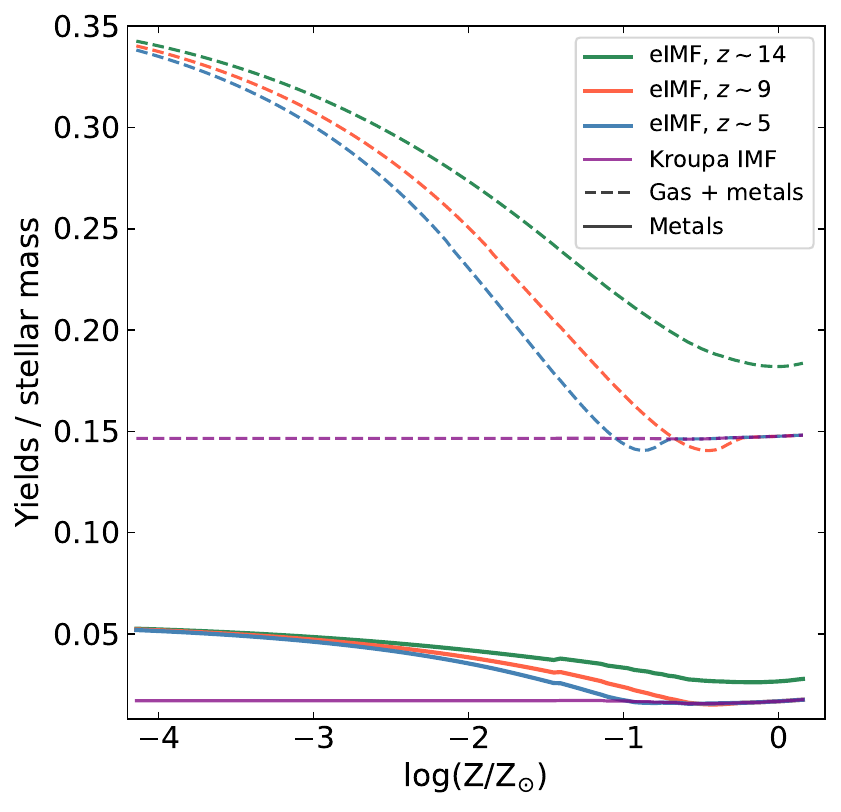}}
  \caption{Evolution of the stellar metal yields (solid lines) and total yields (metals and gas return fraction - dashed lines) per unit stellar mass as a function of metallicity, for the redshifts marked, for the eIMF model. The dashed purple line shows the metallicity-dependent yield values for a Kroupa IMF used in our other models.}
  \label{fig:yields}
\end{center}
\end{figure}

\begin{figure}
\begin{center}
\center{\includegraphics[scale=0.6]{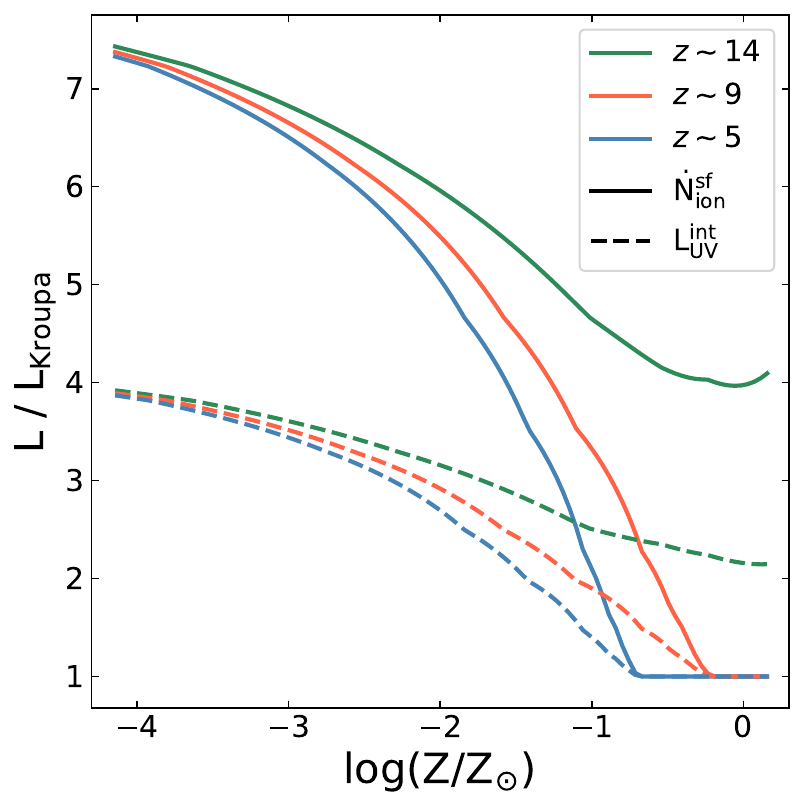}} 
\caption{As a function of metallicity, we show the ratio between the ionising photon production rates (intrinsic UV luminosities) in the eIMF and fiducial model using solid (dashed) lines at the marked redshifts.}
\label{fig:eIMF_lum}
\end{center}
\end{figure}

The first impact of the eIMF model is on the SNII rate ($\nu$), that impacts both feedback and metal/dust production. As shown in Fig. \ref{fig:sn_rate}, the SNII rate has a constant value of $\nu = 0.011 \, \Msun^{-1}$ for the fiducial Kroupa IMF. In the eIMF model, however, the IMF is a function of both redshift and metallicity, which leads to a corresponding change in $\nu$. As shown, the SNII rate increases by about 2.3 times, to $\nu \sim 0.025 \, \Msun^{-1}$ for $Z \sim 10^{-4}\Zsun$, at all $z \sim 5-14$. The value of $\nu$ decreases with increasing metallicity as expected - at $z \sim 5 ~(9)$, the SNII rates in the eIMF and Kroupa IMFs converge for $Z \gsim 0.1 ~ (0.25) \, \Zsun$. However, the SNII rates saturate at a value that is about 1.25 times higher ($\nu \sim 0.014  \Msun^{-1}$) at $z \sim 14$, even for $Z \sim \Zsun$.  

We then calculate the impact of this evolving IMF on stellar (dust and gas-phase metals) yields and the total yields (dust, metals and returned gas) using the grids from \citet{Kobayashi20}, the results of which are shown in Fig. \ref{fig:yields}. For a Kroupa IMF, we find constant dust and metal (total) yield values $\sim 0.017 ~ (0.147)$. In the eIMF case, the dust and metals yields at the lowest metallicities are about $0.05 ~(0.34)$, independent of redshift. These values decrease with increasing metallicity all at $z$; however, at any metallicity value, the yields are higher at higher redshifts due to a more top-heavy IMF (see Eq. \ref{eq:f_massive}). At $z \sim 5~(9)$ the dust and metal yields converge to the Kroupa IMF values at $Z \sim 0.1 ~ (0.25)\Zsun$. At $z \sim 14$, the dust and metal yields saturate at values of about $0.028~(0.184)$ even for $Z = \Zsun$. 

Finally, as expected, the eIMF model also impacts the SEDs for early systems. We compute spectra with {\sc SB99} for a list values of $M_c$ distributed between 0.5 and $100 \Msun$ and four values of metallicities between 0.001 and 0.02. For each halo, we interpolate its spectrum using the closest values in our grid of $M_c$ and metallicities. We compute hydrogen-ionising photon production rates, plus luminosities at nine wavelengths between 1300 $\angstrom$ and 2900 $\angstrom$, and at 3500 $\angstrom$ and 4200 $\angstrom$, to be able to compute beta slopes and Balmer break strengths (see Sections \ref{sec:beta} and \ref{sec:bb}). For illustration we show the impact of the eIMF model on the production rate of ionising photons ($\rm{\dot{N}_{ion}^{sf}}$) and on the intrinsic UV luminosity ($\rm{L_{UV}^{int}}$) in Fig. \ref{fig:eIMF_lum}. At all redshifts, $\rm{\dot{N}_{ion}^{sf}}$ ($\rm{L_{UV}^{int}}$) for the eIMF model is almost a factor 8 (4) larger than for a Kroupa IMF at $Z \sim 10^{-4}\Zsun$. The values slowly decrease with increasing metallicity, until they become equal to that from the Kroupa IMF at $Z \sim 0.15 \, \Zsun$ ($0.5 \, \Zsun$) at $z\sim 5$ (9). For redshifts as high as $z \sim 14$, $\rm{\dot{N}_{ion}^{sf}}$ ($\rm{L_{UV}^{int}}$)  saturate at values about 4 (2) times higher than those from the Kroupa IMF, even at solar metallicity. 

\subsubsection{An evolving star formation efficiency} \label{sec:evol_f}

The second model we explore is termed the evolving star formation efficiency model ('SFE'). Here, both the cold gas fractions and star formation efficiencies evolve as a function of halo mass and redshift. 
We use $f_*(M_h, z)=f_*^{\rm{sphinx}}$ for halos with masses in the range of {\sc sphinx}$^{20}$ halos at the corresponding redshift, as well as for all halos below redshift 7. For more massive halos, at $z\geq14$, we have $f_*(M_h, z) = 0.8$, and in between those redshifts we do a weighted average: $f_*(M_h, z) = (14-z)/7 \cdot f_*^{\rm{sphinx}} + (z-7)/7 \cdot 0.8$. The same holds for $f_{\rm{cold}}(M_h, z)$, but with a value of 0.5 at $z\geq 14$.

We summarise the properties of our fiducial model, our two 'extreme' models, as well as the model from \cite{Mauerhofer23} (labelled '{\sc delphi}23') in Table \ref{table:model_summaries}. We note that the {\sc delphi}23 model did not model cold gas explicitly, but assumed a fraction of $50\%$, as mentioned in Sect. \ref{sec:dust}. 
In the fiducial model presented here, on the other hand, both the cold gas fractions and star formation efficiencies are derived from the {\sc sphinx}$^{20}$ simulations. The only free parameters in the new model are the fraction of SNII energy coupling to gas ($f_w$) and the gas distribution radius ($r_{\rm{gas}}$) that are chosen so as to simultaneously match to the observed evolution of the UV LF and stellar mass function at $z \sim 5-9$, and dust observables at $z \sim 5-7$ from ALMA observations \citep{Bethermin20, REBELS}.  

\begingroup
\setlength{\tabcolsep}{6pt} 
\renewcommand{\arraystretch}{1.3} 
\begin{table}
  \caption{Summary of the four models considered in this work.}
  \label{table:model_summaries}
   \centering
  \begin{tabular}{l|cccc}
    \toprule
    variable  & {\sc delphi}23 & fiducial & eIMF & eSFE \\ 
    \midrule
    $f_*$   & $0.15$ & $f_*^{\rm{sphinx}}$  &  $f_*^{\rm{sphinx}}$ & $f_*(M_h, z)^{\ddag}$ \\
    $\fcold$ & - & $f_{\rm{cold}}^{\rm{sphinx}}$ & $f_{\rm{cold}}^{\rm{sphinx}}$ & $f_{\rm{cold}}(M_h, z)^{\ddag}$ \\
    $f_w$  & 0.06 & 0.04 & 0.04  & 0.04 \\
    $r_{\rm{gas}}$  & $\propto \frac{1+z}{7}$ & $\propto \frac{1+z}{6}$  & $\propto \frac{1+z}{6}$  & $\propto \frac{1+z}{6}$ \\
    IMF & Kroupa & Kroupa & $f(Z,z)^{\dag}$ & Kroupa \\
    \midrule
    $\fesc$ & $13.1 \%$ & $15.2^{+0.5}_{-0.3}\%$ & $5.7^{+0.4}_{-0.2}\%$ & $15.4^{+0.3}_{-0.3}\%$ \\
    \bottomrule
  \end{tabular}
  \tablefoot{The first five rows describe the model parameters: the star formation efficiency ($f_*$), the cold gas fraction ($\fcold$), the coupling between SNII energy and gas ($f_w$), the redshift-dependent factor entering the gas distribution radius ($r_{\rm{gas}}$, see Sect. \ref{sec:def_UV}), and the initial mass function. The sixth row shows the best-fit escape fraction of ionising photons required to match to reionisation observables, as presented in Sect. \ref{sec:ion_emi}. \newline $^{\dag}$ The IMF becomes more top-heavy at low metallicity and high redshift in this model (Sect. \ref{sec:evol_imf}). $^{\ddag}$ See Sect. \ref{sec:evol_f} for details about the mass and redshift dependence.}
\end{table}
\endgroup

\section{Confronting models and observations in the first billion years} 
\label{sec:predictions}

We now discuss the calibration of the model parameters against the latest UV LF estimates obtained from the JWST for all of the models considered in this work. We then compare our models against a number of early galaxy observables including the UV luminosity density, the evolving stellar mass function, the dust masses, the mass-metallicity relation, the beta slopes, Balmer break strengths and, the ionising photon production efficiency. In what follows, we compare the results of the new model against those from the last version of the code referred to as {\sc delphi}23 \citep{Mauerhofer23}.

\subsection{The redshift evolution of the ultraviolet luminosity function at $z \sim 5-20$}
\label{sec:uvlf}

\begin{figure*}
  \resizebox{\hsize}{!}{\includegraphics{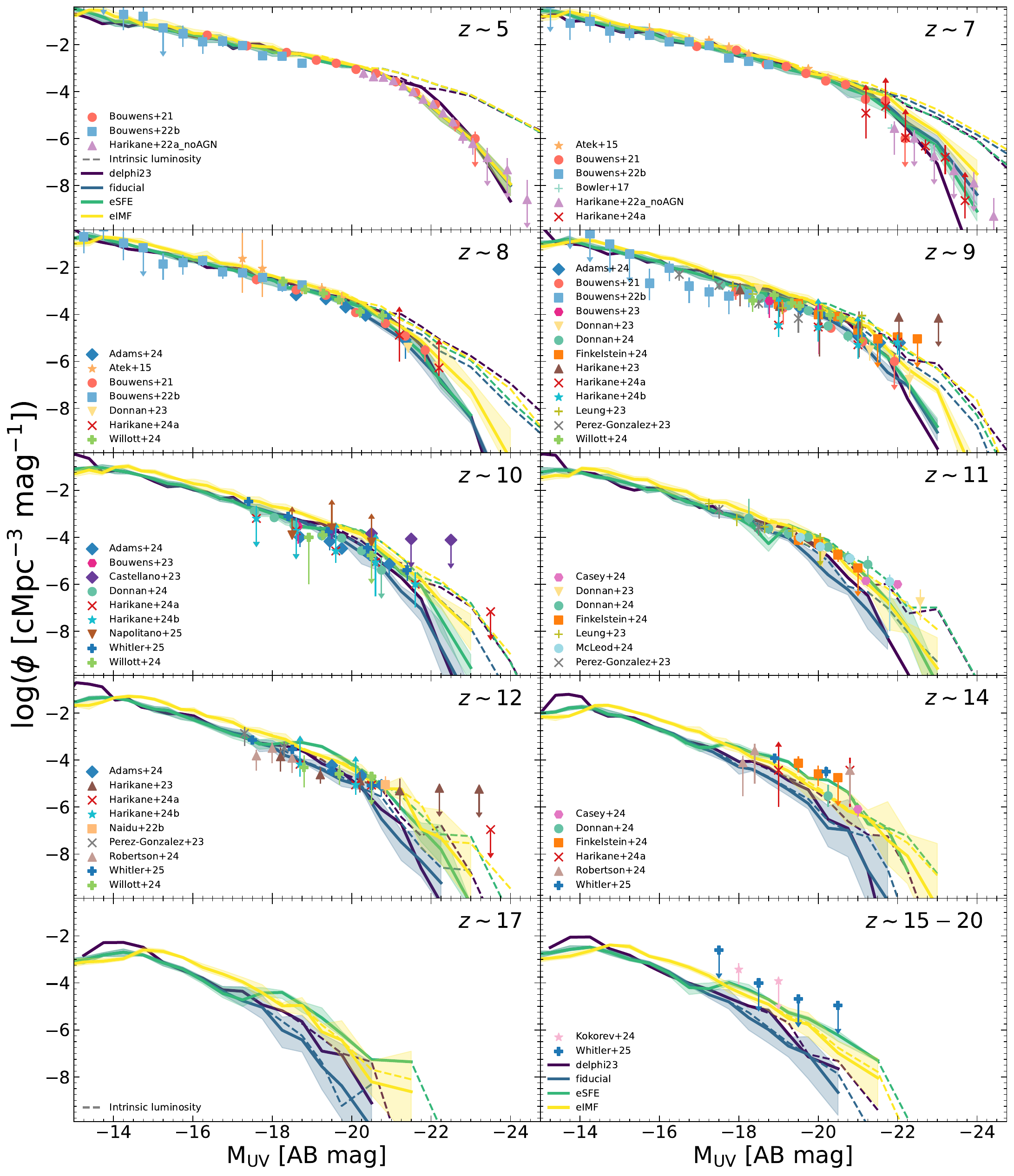}}
  \caption{Redshift evolution of the UV LF between $z \sim 5-20$, as marked. In each panel, we show a comparison of the fiducial model, and the eIMF and eSFE models against the results from {\sc delphi}23 \citep{Mauerhofer23} for both the intrinsic (dashed lines) and dust attenuated (solid lines) UV LFs. These lines show the means of five runs sampling the PDFs for $\fcold$ and $f_*$ from {\sc sphinx}$^{20}$. For the sake of clarity we only show the maximum and minimum values associated with these runs for the observed UV LFs through shaded areas. In each panel, as marked, points show observational data \citep[from][]{Adams24, Atek15, Atek18, Bouwens21b, Bouwens22faint, Bouwens23jwst, Bowler17, Casey24, Castellano23, Donnan23, Donnan24, Finkelstein24, Harikane22a, Harikane23, Harikane24, Harikane24_spec, Ishigaki18, Leung23, McLeod24, Naidu22, Napolitano25, Perez-Gonzales23, Robertson24, Willott24, Whitler25}. The last panel shows the UV LF integrated between $z \sim 15-20$ in order to compare with observations at those redshifts \citep{Kokorev24, Whitler25}. }
  \label{fig:UV_LF_compare}
\end{figure*}

 We now discuss the redshift evolution of both the intrinsic and observed (dust attenuated) UV LFs. In addition to comparing these to available observations at $z \sim 5-14$, we show model predictions out to $z \sim 20$. Starting at $z \sim 5$, we briefly recap that the intrinsic UV LF from the {\sc delphi}23 model over-predicts the bright end of the observed UV LF for $\MUV \lsim -21~ (-22)$ at $z \sim 5 ~(7)$ as shown in Fig. \ref{fig:UV_LF_compare}. Adding dust attenuation results in observed UV LFs in accordance with observations at these redshifts. The importance of dust attenuation decreases with increasing redshift in all of the models considered here. This is driven by gas and dust (that are assumed to be perfectly mixed) occupying an increasing volume of the virial radius, that is getting more 'diffuse' with redshift, leading to a decrease in the dust optical depth. While dust still plays a (decreasing) role in attenuating the UV luminosity out to $z \sim 12$, the intrinsic and observed UV LFs converge at all $z \gsim 14$, meaning that dust attenuation plays no role at these early epochs. In this model, even the intrinsic UV LF under-predicts the bright end of the observed UV LF, with $\MUV \lsim -21$, at $z \sim 12-14$. 

In the fiducial model presented in this work, as noted, the value of $\feff$ (composed of the cold gas fraction and star formation efficiency) are directly obtained from the {\sc sphinx}$^{20}$ simulations. Compared to the {\sc delphi}23 model, we find $\feff$ values that are a factor 5 (10) lower for $\mh \sim 10^{10}\Msun$ halos at $z \sim 5~(17)$. These lower star formation rates are compensated by invoking a slightly weaker coupling between SNII energy and gas ($f_w=0.04$) as compared to the 1.5 times higher value ($f_w=0.06$) used in the {\sc delphi}23 model. As a result, in this new fiducial model, galaxies lose less gas at any time-step, both in terms of star formation and outflows. While the lower star formation rates result in the intrinsic UV LF from the fiducial model being underestimated slightly (by about 0.3-0.5 dex for a given UV magnitude) at the bright end ($\MUV \lsim -20$) at $z\gsim 8$ compared to that from {\sc delphi}23, self-regulation of star formation and feedback results in the intrinsic UV LFs from both these models converging by $z \sim 5$.  
In this fiducial model, the intrinsic UV LF overestimates the bright end ($\MUV \lsim -21$) of the observed UV LF at $z \lsim 8$. Reconciling these requires dust attenuation, the impact of which again decreases (due to a decrease in the dust optical depth) with increasing redshift. Although the bright end of the intrinsic UV LF shows scatter due to the five different random seeds used for $f_*$ and $\fcold$, the mean UV LF under-predicts the bright end of the UV LF for $\MUV \lsim -20.5$ at $z \sim 11-12$, and as faint as $\MUV \sim -19.5$ by $z \sim 14$: for example, it falls short of observations \citep{Casey24,Donnan24, Naidu22, Finkelstein24, Robertson24}  by $\gsim 1$ dex at $z \sim 11-14$.

We then appeal to the eIMF model that produces up to four times higher luminosity per unit star formation as compared to the fiducial model. As shown in the same figure, this model yields an intrinsic UV LF that is in very close agreement with that from {\sc delphi}23, that is it overestimates the observations at all $z \sim 5-14$. Dust is required in order to reconcile this model with observations for $\MUV \lsim -21$ at $z \sim 5-10$; by $z\sim 11$ the intrinsic and observed UV LFs converge due to a decreasing dust attenuation. Interestingly, this model is able to reproduce all JWST observations, including the bright end of the evolving UV LF between $z \sim 5-20$ \citep{Casey24, Donnan24, Finkelstein24, Harikane24, Robertson24,Kokorev24, Whitler25}.  
The other model designed to boost the UV luminosity of {\sc delphi} halos is the 'eSFE' model, where the cold gas fractions and star formation efficiencies of massive halos (above the largest {\sc sphinx}$^{20}$ halo masses) slowly increase with respect to the fiducial model at $z >7$. This model also yields an intrinsic UV LF that over-predicts the data at $z \sim 5-10$ and, including dust, this model successfully reproduces JWST observations of the UV LF, within error bars, at all redshifts $z \sim 5-20$. From $z \sim 10$ to 14, we also see the transition between the mass regime in which $f_*$ is sampled from {\sc sphinx}$^{20}$ and the regime in which it has a redshift-dependent value (see Table \ref{table:model_summaries}). This creates a moderate flattening of the UV LF, yielding a slight discrepancy compared to observations at $z \sim 11-12$ at $\rm M_{UV} = -19$. Given the increasing value of $\feff$, the eSFE model predicts a larger number density of galaxies as compared to the eIMF model for $\MUV \sim -18$ to $-21$ at $z \sim 14-17$; the eIMF model provides the upper limit to the observed UV LF at all other magnitudes for $z \sim 5-21$. Indeed, at the highest redshifts of $z \sim 15-20$, the high SFE values in this model result in the best agreement with observations.

Quantitatively, using the Calzetti extinction curve, we find similar values of dust attenuation E(B-V) of massive galaxies for each model: at $z\sim 5$, galaxies above a stellar mass of $10^{10} \Msun$ have an average and maximum value of E(B-V) of $0.19$ and $0.36$, respectively. As expected, given the decreasing impact of dust with increasing redshift, by $z\sim10$, galaxies above a stellar mass of $10^{9} \Msun$ have an average and maximum E(B-V) value of 0.056 and 0.11.

Finally, we note that while the faint ends of the UV LF are relatively similar for all the models studied here, we see a large spread in predictions at the bright end that can be tested with forthcoming JWST observations.  
\subsection{The redshift evolution of the UV luminosity density at $z \sim 5-21$} \label{sec:sfr_den}

\begin{figure} 
  \resizebox{\hsize}{!}{\includegraphics{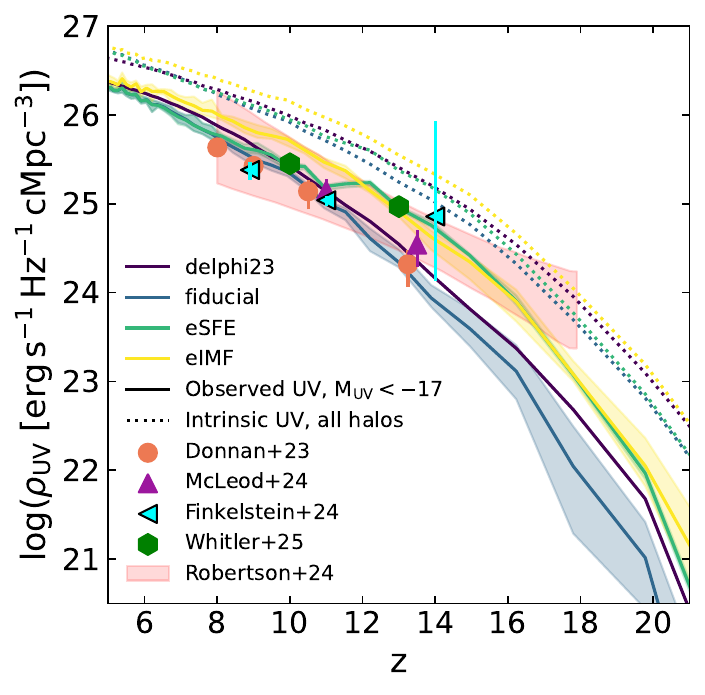}}
  \caption{Redshift evolution of the UV luminosity density at $z \sim 5-21$. Solid lines show the mean of five runs for each model, using dust attenuated values of $\rhouv$ integrating down to observed magnitude limits of $\MUV \lsim -17$. Shaded areas show the maximum and minimum values associated with these runs. Dotted lines show values of the intrinsic luminosity density integrating over all galaxies. As marked, points show observational results \citep{Donnan23, McLeod24, Finkelstein24, Whitler25} and the faint pink shaded area corresponds to the 16\% and 84\% marginal constraints from the JADES origin field \citep{Robertson24}.}
  \label{fig:rho_sfr_compare}
\end{figure}

We now compare the UV luminosity density ($\rhouv$) inferred from our models against observationally inferred results in Fig. \ref{fig:rho_sfr_compare}. We start by discussing the dashed lines, that represent the intrinsic UV luminosity density integrating over all galaxies in our models. The fiducial model shows $\rhouv~ [\rm {erg~ s^{-1} Hz^{-1} cMpc^{-3}}]$ values that decreases by about four orders of magnitude from $\sim 10^{26.7}$ at $z \sim 5$ to $\sim 10^{22.75}~$ by $z \sim 20$. As might be expected from the discussion in Sect. \ref{sec:uvlf} above, while the {\sc delphi}23 model shows $\rhouv$ values higher than the mean fiducial model by about 0.2-0.5 dex at $z \sim 12-20$, these converge at lower redshifts. Producing the largest amount of light per unit star formation, the eIMF model forms the upper limit to the UV luminosity density at all redshifts with $\rhouv \sim 10^{26.8} ~ (10^{23.1})~ [\rm {erg~ s^{-1} Hz^{-1} cMpc^{-3}}]$ at $z \sim 5 ~(20)$; the eSFE model yields $\rhouv$ values that lie between the bounds provided by the fiducial and eIMF models. Interestingly, despite the different star formation prescriptions used, all of these models yield $\rhouv$ values that are different by less than 0.5 dex at any $z \sim 5-21$. 

\begin{figure*}
  \resizebox{\hsize}{!}{\includegraphics{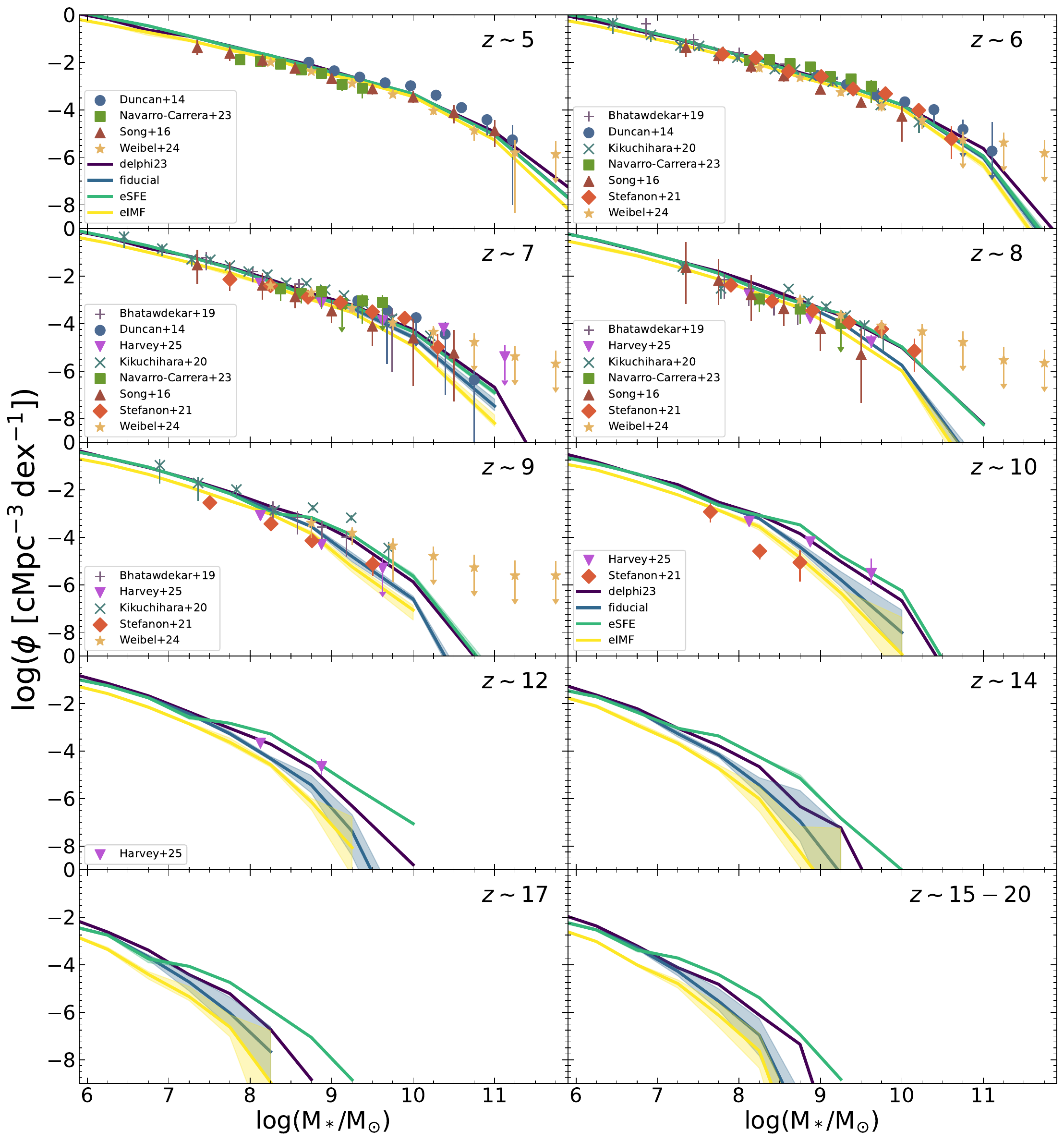}}
  \caption{Redshift evolution of the SMF at $z \sim 5-20$, for the different model explored in this work. Solid lines show the mean results from five runs of each model with shaded areas showing the associated maximum and minimum values. The bottom right panel shows the SMF integrated on all outputs between redshifts 15 and 20. As marked, each panel up to $z\sim12$ show observational data, all assuming a Kroupa IMF \citep[from][]{Bhatawdekar19, Duncan14, Kikuchihara20, Navarro-Carrera24, Song16, Stefanon21, Weibel24, Harvey25}.}
  \label{fig:SMF_compare}
\end{figure*}

We also compute the $\rhouv$ including the impact of dust attenuation and integrating only down to observed limits of $\MUV \sim -17$, the results of which are shown in the same figure as solid lines. The fiducial model shows $\rhouv$ values that decrease from $10^{26.3}$ at $z \sim 5$ to $10^{21}$ by $z \sim 20$, meaning that galaxies with $\MUV \lsim -17$ only contribute about $3.5\%$ of the UV luminosity at $z \sim 20$, increasing to about $40\%$ by $z \sim 5$. Again, the {\sc delphi}23 model yields $\rhouv$ values that are larger by 0.2-0.5 dex at $z \sim 12-20$. With its higher production of UV photons, the eIMF model yield upper limits to the UV luminosity density with $\rhouv \sim 10^{26.4} ~ (10^{22})$ at $z \sim 5~ (20)$; while the eSFE model lies very close to the eIMF model at $z \gsim 13$, at lower redshifts it transitions to the fiducial model as might be expected. 

Given the large errors associated with observations \citep[especially][]{Robertson24}, as of now, all of these models are in reasonable agreement with existing observations at $z \sim 8-14$. The eSFE and fiducial models are in better agreement with the values inferred at $z \sim 8-9$ \citep[from][]{Donnan23,Finkelstein24} and all of the models are in agreement with the data at $z \sim 10-11$ \citep{Donnan23, McLeod24, Finkelstein24, Whitler25}, although the eIMF sits slightly above the data. At $z \sim 13-15$, where the data points from different works show values separated by about 1 dex, the fiducial model is in agreement with the results from \citep{Donnan23} while the eIMF and eSFE models are in better agreement with the higher values inferred by \citet{Finkelstein24} and \citet{Whitler25}. As shown, the model differences increase with increasing redshift, with the eIMF (and eSFE) models predicting $\rhouv$ values an order of magnitude above that from the fiducial model - the slope evolution of the $\rhouv-z$ relation will therefore be an excellent observable to distinguish these models with forthcoming JWST observations.

\subsection{The redshift evolution of the stellar mass function from $z \sim 5-20$}

We also compare the SMF predicted by our models with those constructed from observations with for example the Hubble Space Telescope (HST) and the JWST, as shown in Fig. \ref{fig:SMF_compare}. At $z \sim 5-6$, all four of the models studies in this work show SMFs that essentially overlap between $M_* \sim 10^{6-11}\Msun$; the {\sc delphi}23 model shows a slightly enhanced high-end tail driven by its large star formation efficiency. As we move to $z \gsim 7$, these models start showing increasingly larger differences: with its top-heavy IMF, the eIMF model has the largest light-to-mass ratio in addition to the largest SNII fraction per unit mass. As a result, this sets the lower limit to the theoretical SMF evolution. As seen, this model shows a 1.5 (2.5) times lower amplitude as compared to the fiducial model at $z \sim 9~(17)$. On the other hand, the eSFE model shows the opposite trend. Thanks to its larger effective star formation efficiencies in massive halos above redshift 7, it shows a much larger number of galaxies per unit volume. For example, it shows a factor 10 (100) more galaxies per unit volume at $M_* \sim 10^9 ~\Msun$ at $z \sim 9 ~ (14)$ as compared to the fiducial model.

In terms of observations, the {\sc delphi}23 model is in accordance with the observationally inferred SMF at $z \sim 5-9$. Within error bars, we find the SMF from the fiducial model to also be in accordance with observations at all $z \sim 5-9$. It does, however, slightly underestimate the bright end of the SMF at $z \sim 8$ at $M_* \sim 10^{10.25}\Msun$ \citep{Stefanon21} and the observations from \cite{Kikuchihara20} at $z \sim 8-9$; the latter, however, lie a factor $\sim3$ above observations from other groups at these redshifts \citep{Song16, Bhatawdekar19, Stefanon21, Navarro-Carrera24}. Given its increasing star formation efficiency above $z \sim 7$, the eSFE model sets the upper limit to the SMF. This globally makes it closer to observations, except for \cite{Song16} at $z\sim 8$ and \cite{Stefanon21} at $z\sim 9$. As the lower mass limit, the eIMF model underestimates the bright end of the observed SMF at all $z \sim 8-9$ from, for instance, \citet{Stefanon21} and \citet{Weibel24}, although it is in excellent agreement with the results from \citet{Stefanon21} at $z \sim 9$ at the low-mass end. We caution against a direct comparison of the eIMF model against these observations directly since they all employ a constant Kroupa IMF; assuming an increasingly top-heavy IMF (as in the eIMF) model would naturally result in lower stellar masses being inferred from observations. 
\subsection{The dust-to-stellar mass relations in early systems} \label{sec:gas_dust}

\begin{figure*}
\begin{center}
\center{\includegraphics[scale=0.7]{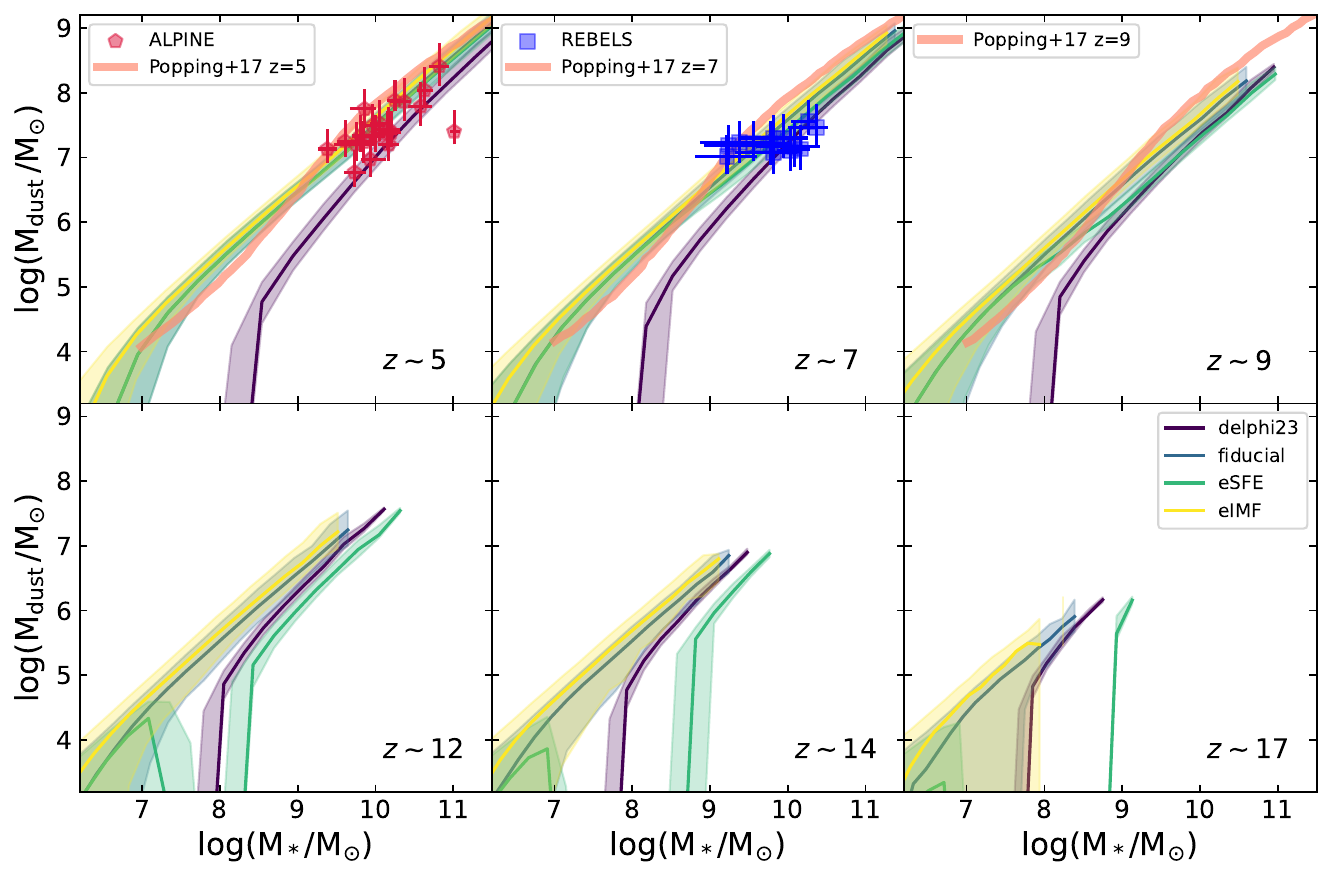}} 
\caption{Evolution of the dust mass as a function of stellar mass at $z \sim 5-17$ for the different models studied in this work, as marked. Lines represent the mean of the five runs with different seeds for each model; shaded areas represent the range between the minima and maxima shown by the \sixteenth and \eightyfourth percentiles, respectively. Points show observational results from the ALPINE survey \citep{Fudamoto20} at $z \sim 5$ and from the REBELS survey \citep{REBELS} at $z \sim 7$, as marked. The thick salmon lines in the first three panels show theoretical results from \cite{Popping17}. }
\label{fig:dust_mass_compare}
\end{center}
\end{figure*}

We now study the redshift evolution of the dust-to-stellar mass relation, for our four models, at $z \sim 5-17$, as shown in Fig. \ref{fig:dust_mass_compare}. In the {\sc delphi}23 model, low-mass galaxies with $\Mstar \lsim 10^{8.5}~ (10^9)\Msun$ at $z \sim 15 ~(7)$ form stars at the 'ejection' efficiency, meaning that any episode of star formation ejects the remaining gas and dust, that are assumed to be perfectly mixed, from the system. This results in low-mass systems bringing in little to no gas and dust into larger successors at later times. However, at every step, the processes of star formation and dust are very closely related (Eqs. \ref{eq:MZ}-\ref{eq:Md}), which results in a linear $M_d-M_*$ relation at any redshift with $M_d \sim 10^{-3.1}M_*$.

In contrast, the other models used in this work that use the {\sc sphinx}$^{20}$ simulations (fiducial, eIMF and eSFE), have lower star formation efficiencies. This results in a situation where even low-mass galaxies do not eject all their gas and dust and can therefore contribute to enriching larger systems at later times. However, the relation between star formation and the key dust processes ensures an almost linear $M_d-M_*$ relation in all models. The key differences between the fiducial and eIMF, and {\sc delphi}23 models are that: firstly, the dust masses have non-zero values down to the lowest stellar masses, as compared to the {\sc delphi}23 model where much larger systems ($\sim M_* \sim 10^{8-8.5}\Msun$) are effectively dust-free due to SNII feedback. Secondly, due to lower amounts of gas and dust being lost in outflows, these relations show shallower slopes for the $M_d-M_*$ relation. For example, these new models show dust masses that are about 0.5 dex higher than the {\sc delphi}23 model for $M_* \sim 10^{9.5}\Msun$ at $z \sim 5-7$; this difference decreases with increasing redshift such that these results converge (within the dispersions for different random seed runs) for all but the lowest mass systems. Interestingly, the eIMF model results in dust masses close to the fiducial model, even though the metal yields are IMF-dependent. This is because the metal yields change significantly only for extremely low metallicity galaxies (see Fig. \ref{fig:yields}); as soon as galaxies get metal-enriched, the yields become similar to the fiducial model. As for the eSFE model, at $z \sim 5-7$, it effectively looks the same as the fiducial model, by construction. However, its behaviour becomes much more complex with increasing redshift. Above $z\sim 12$, we notice three regimes in the dust mass - stellar mass relation of the eSFE model. At masses below $10^{7}\Msun$, galaxies are assigned properties from {\sc sphinx}$^{20}$ halos rendering the relations almost identical to the fiducial model. For $M_* \sim 10^{7-9}\Msun$, however, the SFE is boosted compared to the {\sc sphinx}$^{20}$ values; these galaxies are feedback limited, resulting in them being dust-free. Finally, at even larger masses, while the SFE is still boosted above fiducial values, the host halos are massive enough to hold on to a fraction of gas and dust; this model therefore sets the lower limit to the dust-stellar mass relation at $z \gsim 12$.

In the first two panels of Fig. \ref{fig:dust_mass_compare} we also plot the dust and stellar masses based on ALMA observations of FIR emission, from the ALPINE program at $z\sim 5$ \citep{Fudamoto20} and the REBELS program at $z\sim 7$ \citep{REBELS}. As seen, for the galaxies being observed at $z \sim 5$ with $M_* \sim 10^{9-11}\Msun$, our new models yield dust masses $M_d \sim 10^{7-8.5}\Msun$, or $M_d \sim 10^{-2.4} M_*$, which are in excellent accordance with the ALPINE observations (apart from the most massive system that shows a suppressed dust mass). In terms of REBELS systems, with $M_* \sim 10^{9-10.5}\Msun$, our model yields values of $M_d \sim 10^{6.8-7.5}\Msun$ that are again in excellent accordance with observational results. Finally, with their lower star formation rates (and feedback), these new models are also in accordance with the results from the {\sc santacruz} semi-analytic model \cite{Popping17}. Overall, given their large error bars and the scatter in the data, all of the new models are equally compatible with currently available observations at $z \sim 5-7$. It is worth mentioning two key caveats involved in these observations: firstly, most of these sources are detected in a single ALMA band, requiring the assumption of a dust temperature in order to obtain a dust mass \citep[see discussion in][]{Sommovigo22a}. Furthermore, the assumed star formation history can significantly affect the inferred stellar masses \citep[see e.g.][]{Topping22}. Overall, within dispersions, the new theoretical models explored here do not show any sensible difference in the dust-to-stellar mass relation at $z \sim 5-12$.

\subsection{The mass-metallicity relation and its redshift evolution} \label{sec:metallicity}

\begin{figure*}
\begin{center}
\center{\includegraphics[scale=0.7]{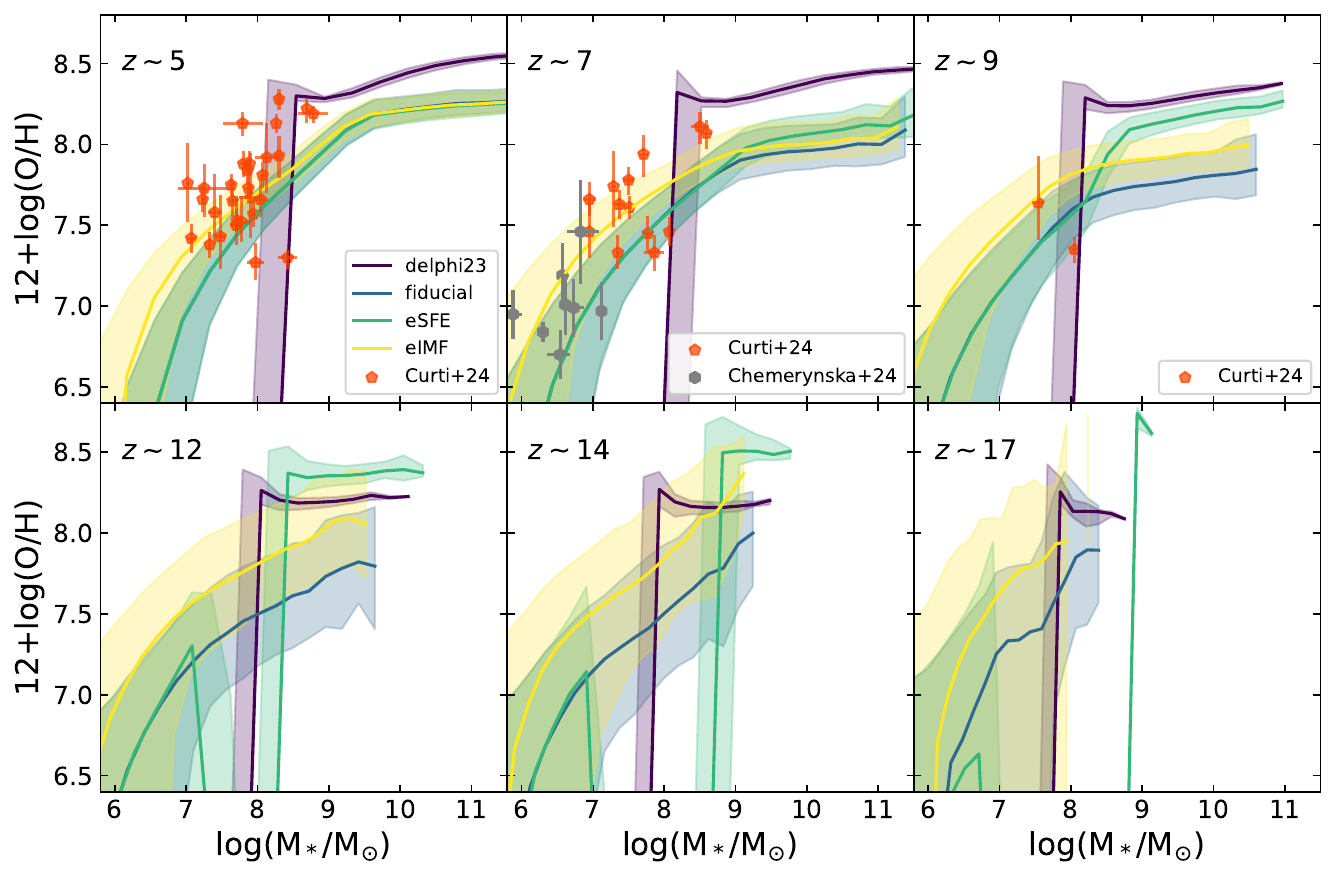}} 
\caption{Gas-phase metallicity as a function of stellar mass at $z \sim 5-17$, as marked. Lines represent the mean of the five runs with different seeds for each model; shaded areas represent the range between the minima and maxima shown by the \sixteenth and \eightyfourth percentiles, respectively. At $z \sim 5-9$, orange pentagons show observational results from \citet{Curti24a}. At $z \sim 7$, grey hexagons show observations of the low-mass end of the mass-metallicity relation from \citet{Chemerynska24}.}
  \label{fig:Z_mass_compare}
\end{center}
\end{figure*}

We now compare the stellar mass-gas-phase metallicity relation of the four models discussed in this work at $z \sim 5-17$, as shown in Fig. \ref{fig:Z_mass_compare}. The metallicity is expressed in standard units in of $12+\log{\rm(O/H)}$ where we use a solar metallicity value of $12+\log(\rm O/H)_{\odot} = 8.69$ \citep{Asplund21}. 

As seen in the previous sections, low-mass galaxies (typically $M_* \lsim 10^{8}\Msun$ at $z \sim 5-9$) lose all of their baryons due to SNII feedback in the {\sc delphi}23 model. As a result, galaxies below these mass limits are devoid of metals, showing a sharp cut-off in the mass-metallicity relation. For larger masses, the metallicity shows a slight increase with increasing mass - for example, at $z \sim 5$ the metallicity value increases from $8.3$ to $8.6$ as $M_*$ increases from $10^{8.5}$ to $10^{11}\Msun$; the overall normalisation of metallicity drops by about $0.2$ dex between $z \sim 5$ and $9$. This model shows the largest metallicity values for high stellar masses at $z \sim 5-9$ given it has the highest values of the star formation efficiency at these redshifts. 

In the new models explored in this work (fiducial, eSFE, eIMF), even systems with masses as low as $M_* \sim 10^6\Msun$ are not feedback limited and show non-zero metallicities; as expected, the metallicity increases as a function of the stellar mass in all of these models. The fiducial model shows a metallicity value of about 7.0 (8.25) for $M_* \sim 10^{7}~ (10^{9.5-11.5})\Msun$ at $z \sim 5$. The normalisation of the mass-metallicity relation decreases slightly with redshift - for example, for a typical stellar mass of $M_* \sim 10^9 \Msun$, the gas-phase metallicity has values of about $8.0, 7.75$, and $7.65$ at $z \sim 5, 9$ and $12$. 

In the eIMF model, at $z \sim 5-7$, low-mass systems ($M_* \lsim 10^8\Msun$) show slightly elevated values of the mean metallicity (by about 0.3 dex) given their top-heavy IMFs; they converge to the fiducial model for more massive systems. Given an increasingly top-heavy IMF (and the associated larger metal yields), this model shows an increase in the mean amplitude of the mass-metallicity relation by about $0.2~(0.4)$ dex at $z \sim 9~(12)$, although the slopes remain similar. At $z \sim 12$, for example, systems with $M_* \sim 10^{9}\Msun$ show metallicity values of about $8.05$.      

The eSFE model shows a similar behaviour to that discussed in the previous section - at $z \sim  5-7$, this model yields results in good agreement with the fiducial model. However, given its cold gas fractions and star formation efficiencies that increase with increasing redshifts, high-mass systems ($M_* \sim 10^{9}\Msun$) at $z \gsim 12$ show the maximum metallicity values in this model; indeed, by $z \sim 9$, massive systems ($M_* \gsim 10^8\Msun$) show about 0.5 dex higher metallicities in this model compared to the fiducial one. Quantitatively, for a system with $M_* \sim 10^{9}\Msun$, the metallicity value increases from about $8.1$ at $z \sim 9$ to $8.4$ by $z \sim 12$ in this model. Further, while galaxies with $M_* \sim 10^{7.5-8} \Msun$ are completely feedback limited, lower mass galaxies still contain gas due to the low star formation efficiency values from {\sc sphinx}$^{20}$ at $z \gsim 12$. 

We also compare our results with recent observational results from the JWST at $z \sim 5-9$ \citep{Curti24a, Chemerynska24}, as shown in the same figure. At $z \sim 5$, within the dispersions associated with the models, the observational values from  \citep{Curti24a} cover the parameter space allowed by all of the models roughly equally well. However, possibly due to our assumption of the instantaneous recycling approximation, we are unable to reproduce the high metallicity values ($12+\log(\rm O/H)>8$) inferred observationally for sources with $M_* \sim 10^{7.5-8.25}\Msun$ \citep[see discussion in Sect. 4.1;][]{Ucci2023}. The situation is similar at $z \sim 7$ where, within the errors and dispersions associated with the data and theoretical models, respectively, the mass-metallicity relation points from both \citet{Curti24a} and \citet{Chemerynska24} agree equally well with all of the new models. Overall here too, within dispersions, the new theoretical models explored do not show any sensible differences in the mass-metallicity relation or its redshift evolution at $z \gsim 5$.

\subsection{Beta slopes for early systems} \label{sec:beta}

\begin{figure*}
\begin{center}
\center{\includegraphics[scale=0.7]{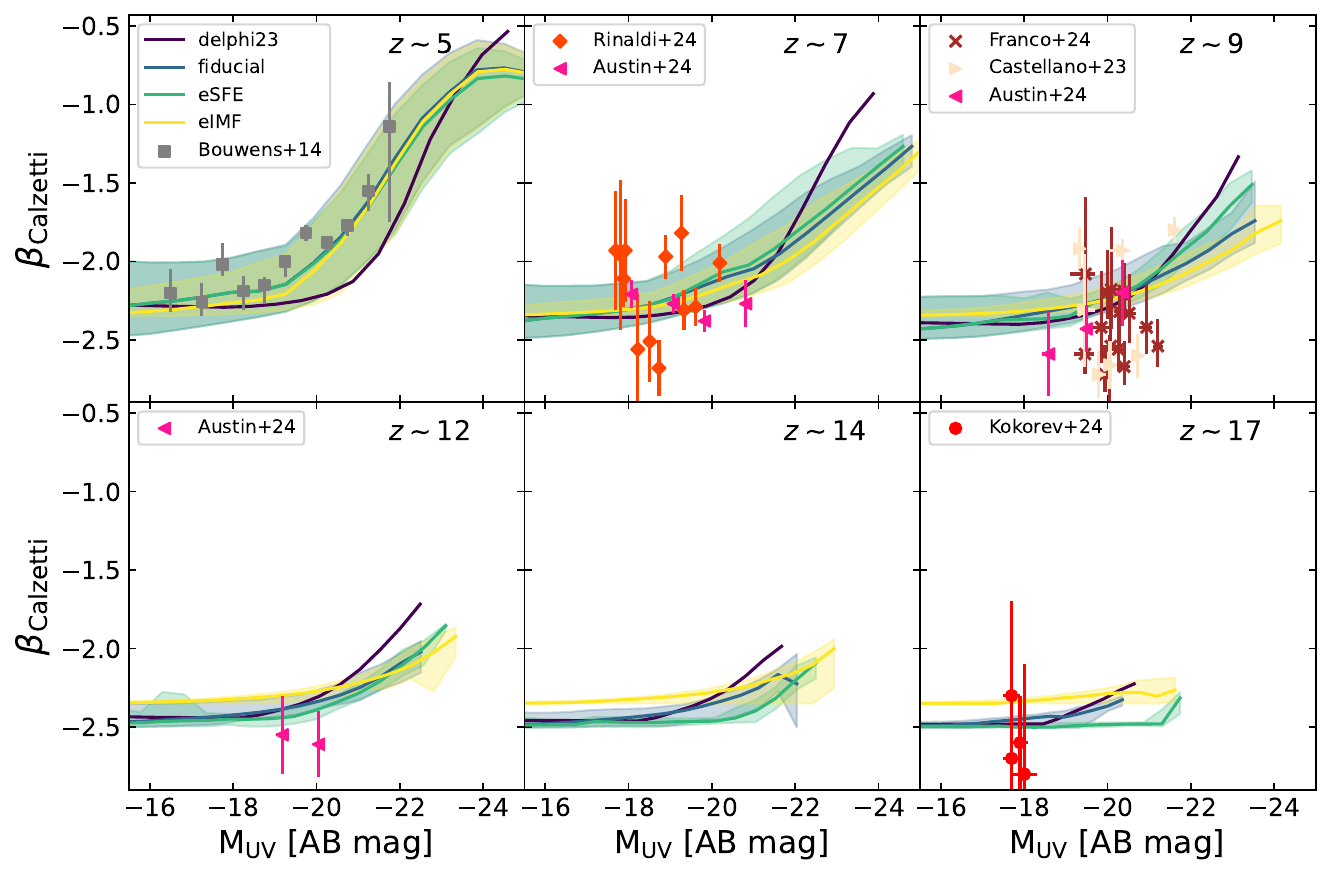}} 
\caption{Ultraviolet spectral slopes ($\beta$), using the Calzetti dust extinction law, as a function of the observed UV magnitude at $z \sim 5-17$, as marked. Lines represent the mean of the five runs with different seeds for each model; shaded areas represent the range between the minima and maxima shown by the \sixteenth and \eightyfourth percentiles, respectively. Grey squares in the top left panel represent observational data from 677 galaxies at $z\sim 5$, taken from \cite{Bouwens14}. As marked in other panels, we also show observational estimates of the $\beta$ slope from recent JWST studies at $z \sim 7-17$ \citep{Castellano23, Austin24, Franco24, Kokorev24, Rinaldi24}.}
  \label{fig:beta_mass_compare}
\end{center}
\end{figure*}

We now show the relation between the observed (dust attenuated) UV spectral slope ($\beta$) and the observed UV magnitude, also called the $\beta-\MUV$ relation, in Fig. \ref{fig:beta_mass_compare}. The intrinsic spectra for each galaxy is convolved with the Calzetti dust extinction law \citep{calzetti2000} using the UV dust attenuation values obtained in Sect. \ref{sec:def_UV} - we then obtain the $\beta$ value by fitting a power-law at nine (equally spaced) wavelengths between $1300-2900\AA$ in the rest-frame of the galaxy. We start by noting that the intrinsic (i.e. dust-unattenuated) UV spectral slope is rather insensitive to the UV magnitude and all four models show values of $\beta \sim -2.3 - -2.5$ at $z \sim 5-7$ (not shown in the figure, for simplicity) as might be expected for stellar populations that are a few tens of Myr old with metallicity values of a few tenths of the solar value. The observed slopes, on the other hand, become 'redder' with an increase in the dust enrichment and/or average stellar ages, meaning that the more massive or luminous a system, the redder is the $\beta$ slope, for all models. As seen from Sect. \ref{sec:gas_dust}, the {\sc delphi}23 model has the lowest amplitude of the dust mass-stellar mass relation for low-to-intermediate mass objects at $z \sim 5-7$ driven by its different prescriptions for the gas available for star formation and the star formation efficiency. As a result, low-to-intermediate luminosity galaxies in this model show the bluest slopes with $\beta \sim -2.3$ at $\MUV \sim -20$. Given their different assembly histories and overall larger dust masses, the new models in this work (fiducial, eIMF and eSFE) show redder slopes with $\beta \sim -2 - -2.3$ at $\MUV \sim -20$; the results from these three models start converging at $z \lsim 9$ as might be expected. At $z \sim 5-9$, the most massive systems in the {\sc delphi}23 model show the reddest $\beta$ slopes as a result of their older stellar populations convolved with high dust masses. At $z \gsim 12$ the $\beta$ slopes from all models start converging to values of about $-2.4$ at $\MUV \sim -20$ as a result of younger stellar populations and low dust attenuations. Even for the most massive/luminous systems, the $\beta$ values from the new models are bluer than $-2$. 

Comparing our results to observations, we find the new models to yield a $\beta-\MUV$ relation that is in excellent agreement with both the slope and amplitude of the (binned) results from \citep{Bouwens14} at $z \sim 5$. Within the current observational error bars, essentially all models agree with results from recent JWST observations at $z \sim 7-17$ \citep[from][]{Castellano23, Austin24, Franco24, Kokorev24, Rinaldi24}; the confirmation of extremely blue systems (with $\beta \ll -2.5$) would pose a challenge to all of the models discussed here. Further, we note that a caveat that enters these comparisons is that the observed $\beta$ slopes, derived from photometry, are affected both by emission lines and nebular continuum, both of which are currently not included in our modelling.

\subsection{Balmer breaks in the first billion years} \label{sec:bb}

\begin{figure*}
\begin{center}
\center{\includegraphics[scale=0.7]{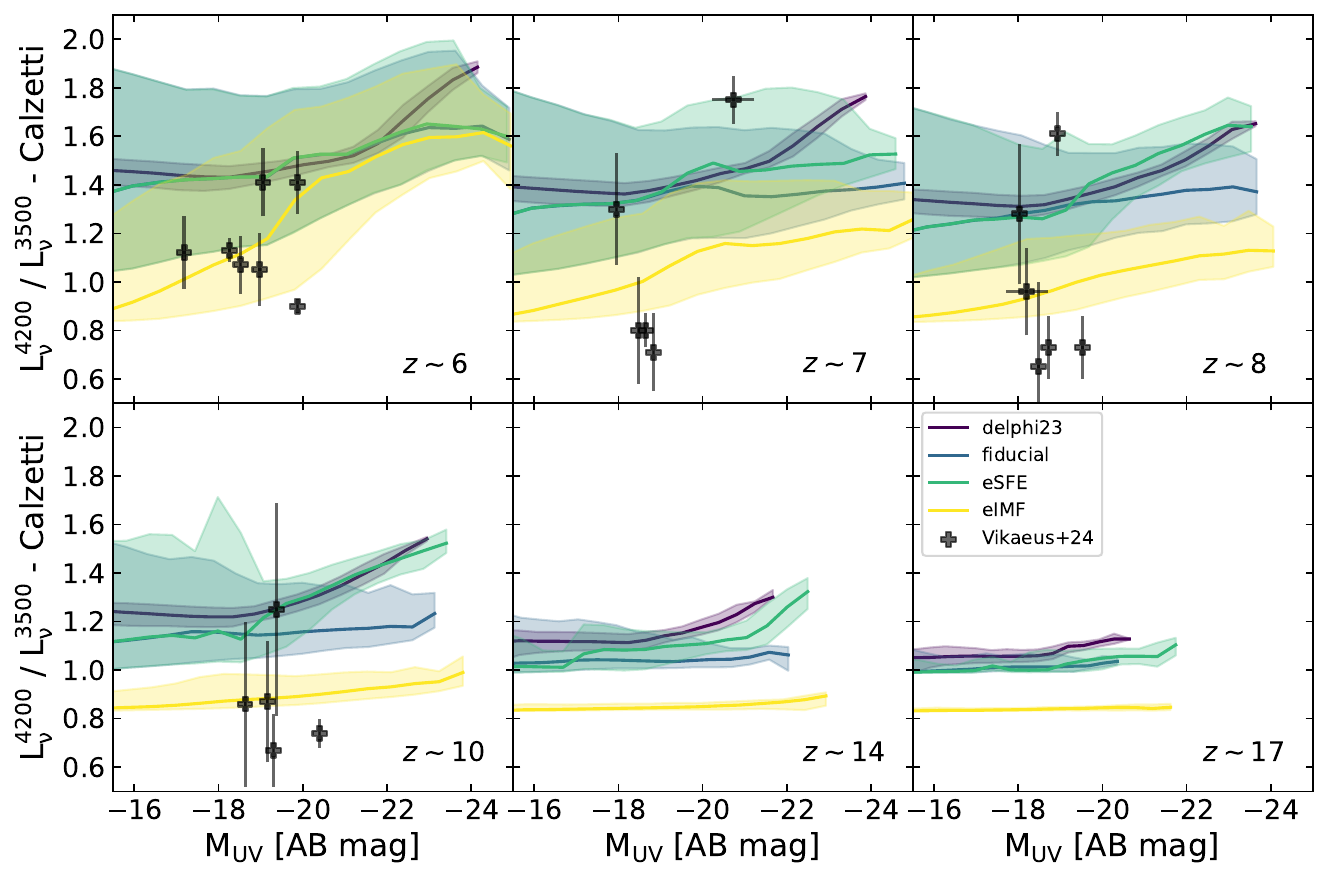}} 
 \caption{Balmer break strength as a function of the observed UV magnitude at $z \sim 6-17$, as marked. Lines represent the mean of the five runs with different seeds for each model; shaded areas represent the range between the minima and maxima shown by the \sixteenth and \eightyfourth percentiles, respectively. Points show observational results from \citet{Vikaeus24}.}
  \label{fig:bb_MUV_compare}
\end{center}
\end{figure*}

We now show the Balmer break strength (the ratio of luminosities at $4200$ and $3500 \AA$) as a function of the observed UV magnitude in Fig. \ref{fig:bb_MUV_compare}. The Balmer break strength depends on the age of stellar populations (with younger populations having weaker breaks), the dust attenuation and the IMF (with more massive stars showing weaker breaks). To compare our models with observations, we compute the Balmer break of all our galaxies assuming the Calzetti attenuation law. At $z \sim 5$, the Balmer break strength increases with increasing luminosity in the eIMF model, from $\lsim 1$ for low-luminosity systems ($\MUV \gsim -18$) to $\sim 1.6$ for the brightest systems given the larger fraction of low-mass stars in the former; for the largest systems, the trend is very similar to that shown by the other three models considered. As the IMF becomes increasingly top heavy with increasing redshifts, the Balmer break strength-$\MUV$ relation shows a progressive flattening with increasing redshift. Indeed, the Balmer break strength saturates to a value of about $0.8-0.9$ at $ \gsim 10$. Given their similar IMFs, the fiducial and eSFE models show very similar values of the Balmer break strength at $z \lsim 8$; while the mean values diverge for the most luminous systems, due to the different assembly histories, these differences lie within the dispersions of these two models. Finally, related to its reddest $\beta$ slopes for luminous systems, the {\sc delphi}23 model shows the largest Balmer break value for these systems. By $z \sim 14$, however, the results from the fiducial, eSFE and {\sc delphi}23 models converge to values $\sim 1.1$, about $20\%$ higher than those from the eIMF model. 

We also compare our results with recent JWST observations presented in \citep{Vikaeus24} as shown in the same figure. Given the dispersions both in the observed Balmer break strengths for a given magnitude and the different models used here, as of now, it is hard to infer a particular model being preferred. One might have a slight preference for the eIMF model to explain the very low values of the strength ($\lsim 0.9$) observed at all $z \sim 6-10$. We again stress the lack of any nebular continuum in our modelling, which can significantly impact the inferred value of the Balmer Break \citep{Katz24_balmer}. Future datasets will be crucial in building larger statistics to shed light on the stellar populations using this statistic.

\subsection{The redshift evolution of the ionising photon production efficiency} \label{sec:xi_ion}

\begin{figure*}
\begin{center}
\center{\includegraphics[scale=0.7]{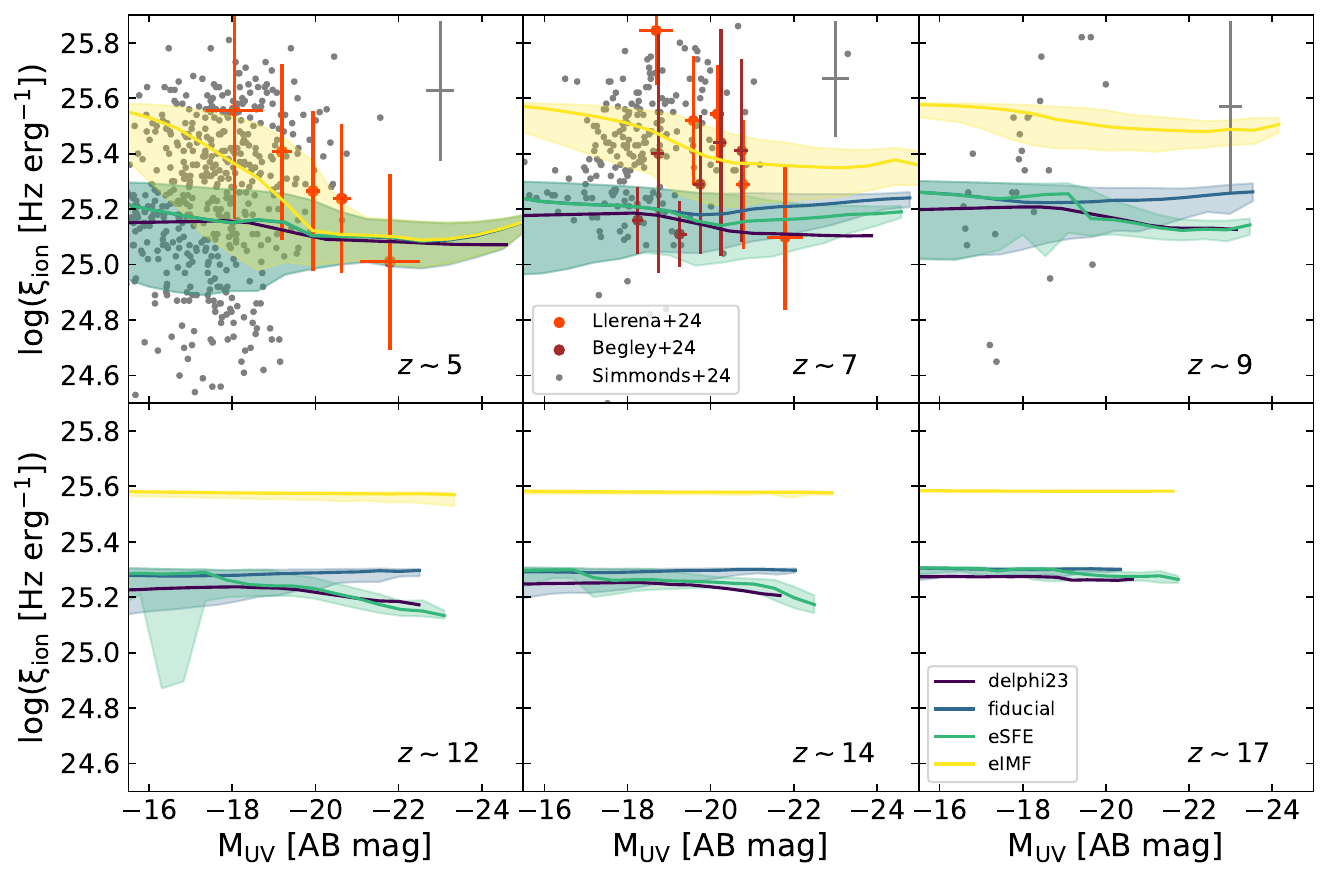}} 
 \caption{Ionising photon production efficiency ($\xion$) as a function of the observed UV magnitude. Lines represent the mean of the five runs with different seeds for each model; shaded areas represent the range between the minima and maxima shown by the \sixteenth and \eightyfourth percentiles, respectively. Points show observational data as marked \citep[from][]{Simmonds24, Llerena24, Begley24}. The crosses at the top right corner of the upper panels indicate the average error on measured magnitudes and $\xion$ as estimated by \cite{Simmonds24}.} 
  \label{fig:xi_ion_compare}
\end{center}
\end{figure*}

A common way of studying the ability of galaxies to contribute to reionisation is through the ionising photon production efficiency, $\xion = \nionsf/L_{UV}^{int}$ where $\nionsf~{\rm[s^{-1}]}$ is the intrinsic production rate of ionising photons from star formation and ${\rm L_{UV}^{int}} ~{\rm[erg~s^{-1}~ Hz^{-1}]} $ is the intrinsic UV luminosity. This quantity strongly depends on the properties of the underlying stellar population including the IMF and the mean stellar age. We show $\xion$ as a function of UV magnitude for $z \sim 5-17$ in Fig. \ref{fig:xi_ion_compare}. As might be expected, the fiducial, eSFE and {\sc delphi}23 models yield very similar results for this quantity given they all use the same IMF. At $z \sim 5-9$, in these models, we find mean values of $\log(\xion) \sim 25.1-25.2~{\rm [Hz~erg^{-1}]}$. However, producing many more massive stars, galaxies in the eIMF model show a much higher value of $\xion$: for example, at $z \sim 5$ faint galaxies with $\MUV \gsim -17$ show values as high as $\log(\xion)=25.55~{\rm [Hz~erg^{-1}]}$. $\xion$ decreases with increasing brightness and metallicities and converges to all the other models by $\MUV \sim -21$. With increasing redshifts, even the brightest systems have flatter IMFs resulting in $\xion$ flattening with UV magnitude. Crucially, in the eIMF model galaxies produce $2-2.5\times$ as many ionising photons per unit UV luminosity as compared to all the other models at $z \sim 12-17$. 

We also compare these results with recent JWST observations of the $\xion-\MUV$ relation at $z \sim 5-9$ \citep[from][]{Begley24, Llerena24, Simmonds24} as shown in Fig. \ref{fig:xi_ion_compare}.  At $z \sim 5-9$, the observations from \citet{Simmonds24} show $\log(\xion)$ values ranging between $24.5-25.8$ for systems with $\MUV \sim -16--20$ and do not show any specific trend as a function of the UV magnitude. This is in accordance with the results from \citet{Begley24} at $z \sim 7$ that range between $\log(\xion) \sim 25-25.8$ for $\MUV \sim -18--20$ and also do not show any specific trend with the UV magnitude - both of these observations are compatible with all of the four models considered here. On the other hand, at $z \sim 5-7$, the average results from \citet{Llerena24} clearly show a decreasing $\xion$ with increasing brightness - for example, their average values of $\log(\xion)$ decrease from about 25.6 to 25 as $\MUV$ decreases from $\sim -18$ to $\sim -22$ at $z \sim 5$. Interestingly, this decrease is in excellent accordance with the eIMF model, where low-mass systems produce copious amounts of ionising photons, at $z \sim 5$ and reasonably compatible at $z \sim 7$. A number of other recent works have also obtained constraints on $\xion$: \citet{atek24} find $\log(\xion) = 25.8\pm 0.14$ for galaxies with $\MUV \gsim -16.5$ at $z \sim 6-7.7$, which is a factor four higher than older works \citep[e.g.][]{robertson2013} and a factor 1.6 higher than the values from any of our models. Further, \citet{Rinaldi24} find $\log(\xion)=25.50^{+0.10}_{-0.12}$ for H$\alpha$ emitters at $z \sim 7-8$, which is in good agreement with results from the eIMF model. Finally, \citet{curtis-lake23} derive $\log(\xion)=25.04^{+0.17}_{-0.21}$ for a source with $M_* \sim 10^{7.58}\Msun$ at $z \sim 10.38$ and $\log(\xion)=24.73^{+0.25}_{-0.15}$ for another object with with $M_* \sim 10^{8.67}\Msun$ at $z \sim 11.58$. These values are lower than our eIMF model by a factor 3.5 and are in much better agreement with the fiducial and eSFE models. Overall, given the variation in observational results, as of now, they do not favour any specific model presented in this work apart from the observations by \citet{Llerena24} that seem to show a preference for the eIMF model.

\section{The epoch of reionisation in light of JWST data} \label{sec:reionisation}

We show the intrinsic and escaping ionising photon emissivity in all of the models presented in this work before discussing the key reionisation sources, and its progress, in this section.

\subsection{The intrinsic and escaping ionising photon emissivities} \label{sec:ion_emi}

\begin{figure*}
\begin{center}
\center{\includegraphics[scale=0.8]{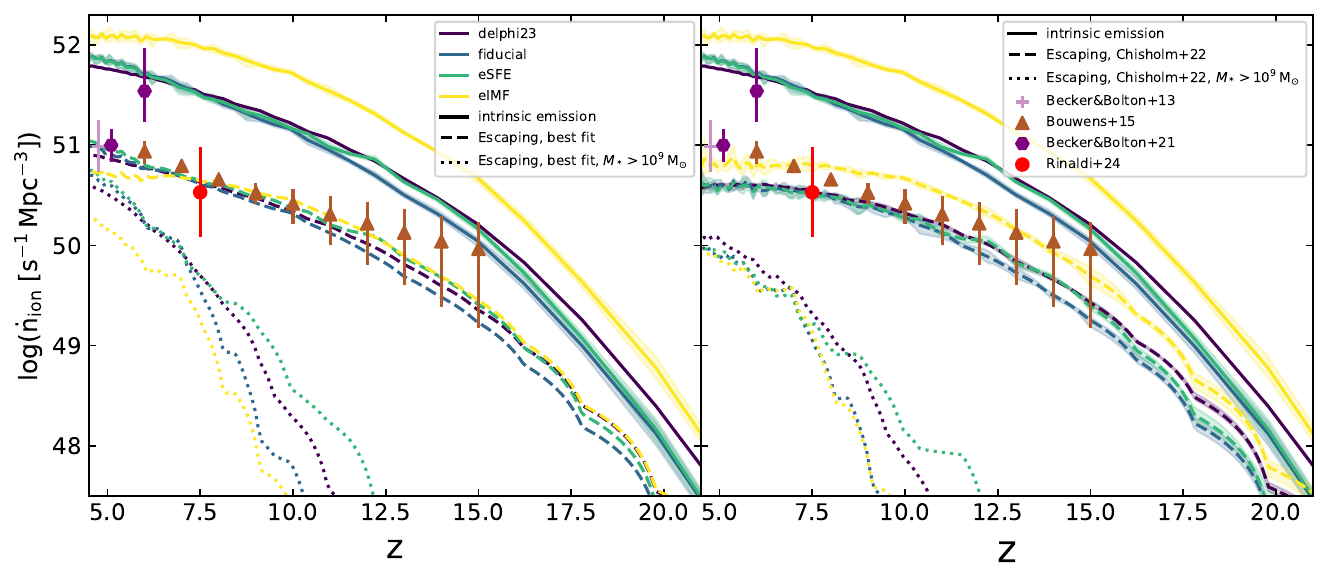}}
\caption{Evolution of the ionising photon rate density as a function of redshift. In both panels, for reference, the solid lines represent the intrinsic ionising photon density for the different models studied, as marked. Shaded regions highlight the spread from the minimum to maximum values among the five runs with different random seeds. The two panels show the 'escaping' emissivity for the two $\fesc$ models studied in this work for all galaxies (dashed lines) and those with $M_*>10^9\Msun$ (dotted lines). The {\it left panel} shows results for a constant value of $\fesc$ needed to match reionisation constraints; the {\it right panel} shows results using $\fesc$ values from \citet{Chisholm22} as shown in Eq. \ref{eq:chisholm_fesc}. In both panels, points show observational constraints \citep[from][]{becker2013, Bouwens15b, becker2021, Rinaldi24}. }
  \label{fig:rho_ion_compare}
  \end{center}
\end{figure*}

We start by discussing the intrinsic emissivity of hydrogen ionising photons, weighted over all galaxies, at $z \sim 5-21$, as shown by the solid lines in Fig. \ref{fig:rho_ion_compare}. The fiducial and eSFE models show results in good accordance with each other, with the emissivity increasing from about $10^{47.7}$ at $z\sim 21$ to $10^{51.8} ~[{\rm s^{-1} Mpc^{-3}}]$ by $z \sim 5$; given its slightly different star formation efficiency, the {\sc delphi}23 model shows slightly higher values (by about 0.2 dex) at $z\gsim 14$, converging to the other models at lower redshifts. In contrast, the eIMF model produces a much larger emissivity than the other models at all redshifts, thanks to its flatter IMF slope. Driven by a combination of high redshifts and low metallicities, it results in an emissivity value that is about $4 \times$ larger than all the other models at $z \gsim 8$; the values are a factor $\sim1.8$ higher by $z \sim 5$, driven by low-mass ($M_*\lsim 10^9 \Msun$), low-metallicity systems.  

We also calculate the escaping emissivity using two models for the escape fraction: {\it (i)}: in the first, we assume a value of $\fesc$ that is constant for all galaxies at all redshifts. This value is obtained by matching to the observed evolution of the volume filling fraction of neutral hydrogen as detailed in what follows; {\it (ii)}: our second model is driven by results from the Low-redshift Lyman Continuum Survey \citep[LzLCS; PI: Jaskot, HST Project ID: 15626,][]{Flury22}. From these observations, $\fesc$ depends on the $\beta$ slope as \citep{Chisholm22}
\be \label{eq:chisholm_fesc}
\fesc^{\rm{chisholm}} = (1.3 \pm 0.6) \times 10^{-4} \times 10^{(-1.22 \pm 0.1)\beta}.
\ee
We compute the $\fesc$ value for each galaxy in each model considered using the $\beta$ values computed in Sect. \ref{sec:beta}. 

A complication is introduced by the fact that the process of reionisation generates a heating UV background (UVB) that can suppress the baryonic content of low-mass halos in ionised regions \citep[e.g.][]{gnedin2000, sobacchi2013a, Hutter21}. To model this feedback effect, we run our models assuming a complete suppression of the gas mass (and hence star formation) in all halos below $M_h = 10^9\Msun$ in ionised regions. The total 'emerging' ionising emissivity for reionisation is obtained by weighing over the UV-suppressed contribution from low-mass halos in ionised regions as well as that from unsuppressed sources in neutral regions, such that \citep{dayal2024}
\be \label{eq:effective_nion}
\rm \dot{n}_{ion} = \sum \phi_{i} \left( Q_{II} \dot{N}_{ion,II}^{sf}(i) \, f_{esc,II}(i) + Q_I \dot{N}_{ion,I}^{sf}(i) f_{esc,I}(i) \right).
\ee
Here, $Q_{II}$ and $Q_I$ represent the volume filling fractions of ionised and neutral hydrogen, respectively. Further, $\rm \phi_i$ is the number density associated with any halo and the first and second terms on the right-hand side show the contribution from suppressed sources in ionised regions and that from unsuppressed sources in neutral regions, respectively. 

The evolution of $Q_{II}$ can be calculated as \citep{Madau17}
\be \label{eq:madau17}
\frac{\mathrm{d}Q_{II}}{\mathrm{d}t} = \frac{\rm\dot{n}_{\rm ion}}{\langle n_{\rm H} \rangle \left(1 + \left\langle \kappa_{\nu_L}^{\rm LLS} \right\rangle / \left\langle \kappa_{\nu_L}^{\rm IGM} \right\rangle \right)} - \frac{Q_{II}}{\bar{t}_{\rm rec}}.
\ee
Here, $\langle n_{\rm H} \rangle$ is the mean hydrogen density of the Universe. The terms $\left\langle \kappa_{\nu_L}^{\rm LLS} \right\rangle$ and $\left\langle \kappa_{\nu_L}^{\rm IGM} \right\rangle$ are the volume averaged absorption probabilities per unit length due to Lyman limit systems and the uniform IGM, respectively, and are given by equations 11 and 13 of \cite{Madau17}. Finally, $\bar{t}_{\rm rec}$ is the effective recombination timescale in the IGM, and is equal to $\left[ (1+\chi) \langle n_{\rm H}\rangle \alpha_0 C_R \right]^{-1}$, where $\chi=0.083$ accounts for the presence of electrons from $\heii$, $\alpha_0$ is the case-A recombination coefficient assuming $T=10^{4.3}$K and $C_R=2.9 \left[(1+z)/6\right]^{-1.1}$ is the clumping factor \citep{Shull12}. We solve Equations \ref{eq:effective_nion} and \ref{eq:madau17} jointly using a backwards differentiation formula method with \textsc{Scipy}'s \texttt{OdeSolver}\footnote{\label{fn:scipy}\url{https://scipy.org/}}, as in \cite{Trebitsch22}.

We show the resulting escaping emissivities of ionising photons as dashed lines also in Fig. \ref{fig:rho_ion_compare} for both $\fesc$ models considered in this work. As discussed in what follows, in the constant $\fesc$ scenario, matching to the redshift evolution of $Q_I$ requires similar average values of $\fesc \sim 15.2\%$ and $15.4\%$ for the fiducial and eSFE models, respectively.  With its larger star formation efficiency value, the {\sc delphi}23 model requires a slightly smaller value of $\fesc\sim 13.1\%$. However, as discussed above, at almost all $z \sim 6-21$, galaxies in the eIMF model produce about three times as many ionising photons per unit volume as compared to the fiducial model; compensating for this requires a much lower value of $\fesc \sim 5.7\%$ in this model in order to match the observed constraints. With these values, we obtain essentially similar evolutions of the escaping emissivities in all four models, which match observed emissivities \citep[from][]{becker2013, Bouwens15b, becker2021, Rinaldi24}. Further, low-mass systems ($M_*\lsim 10^9\Msun$) provide $\sim 85\%$ of ionising photons down to the midpoint of reionisation at $z \sim 7$, dominating the reionisation budget. The contribution of higher mass systems increases with decreasing redshift from about $15\%$ at $z \sim 7$ to about $30\%$ by the end of reionisation, as shown in Fig. \ref{fig:rho_ion_compare}. In the model considering $\fesc$ values using the \citet{Chisholm22} relations, the escaping emissivities again increase with decreasing redshift. While the fiducial, eSFE and {\sc delphi}23 models show very similar results, the overall higher $\xion$ values in the eIMF model lead to a three times higher escaping emissivity at any redshift. Given their larger dust contents (i.e. redder $\beta$ slopes), in this case too, high-mass systems ($\Mstar>10^9\Msun$) have a sub-dominant contribution to the escaping emissivity and contribute at most $15\%$ to the process at any $z \sim 5-21$. Hence, in our model, low-mass galaxies are the key drivers of the reionisation process, even accounting for the impact of reionisation feedback. 

\subsection{The key sources and progress of reionisation} \label{sec:reio_sources}

\begin{figure}
\begin{center}
\center{\includegraphics[scale=0.85]{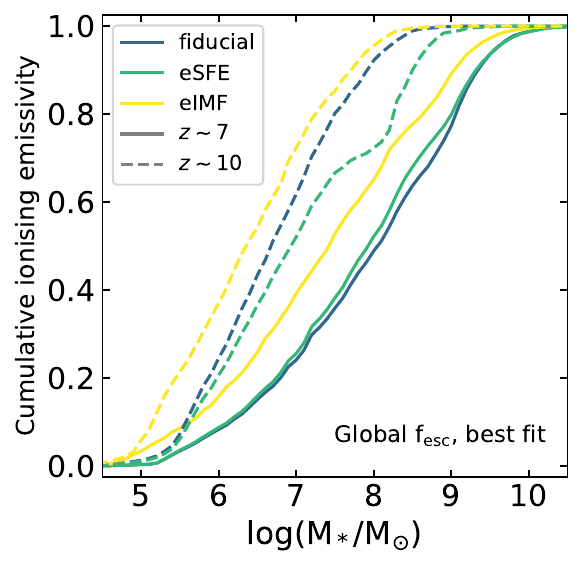}} 
\caption{Cumulative emissivity of 'escaping' ionising photons as a function of stellar mass for the new models explored in this work (fiducial, eIMF and eSFE), as marked. Solid and dashed lines show results at $z \sim 7$ and 10, respectively. For the sake of clarity, we limit results to our constant 
$\fesc$ models, matched to reionisation constraints (see Sect. \ref{sec:reionisation}).}
  \label{fig:cumul_nion}
\end{center}
\end{figure}

\begin{figure*}
\begin{center}
\center{\includegraphics[scale=0.8]{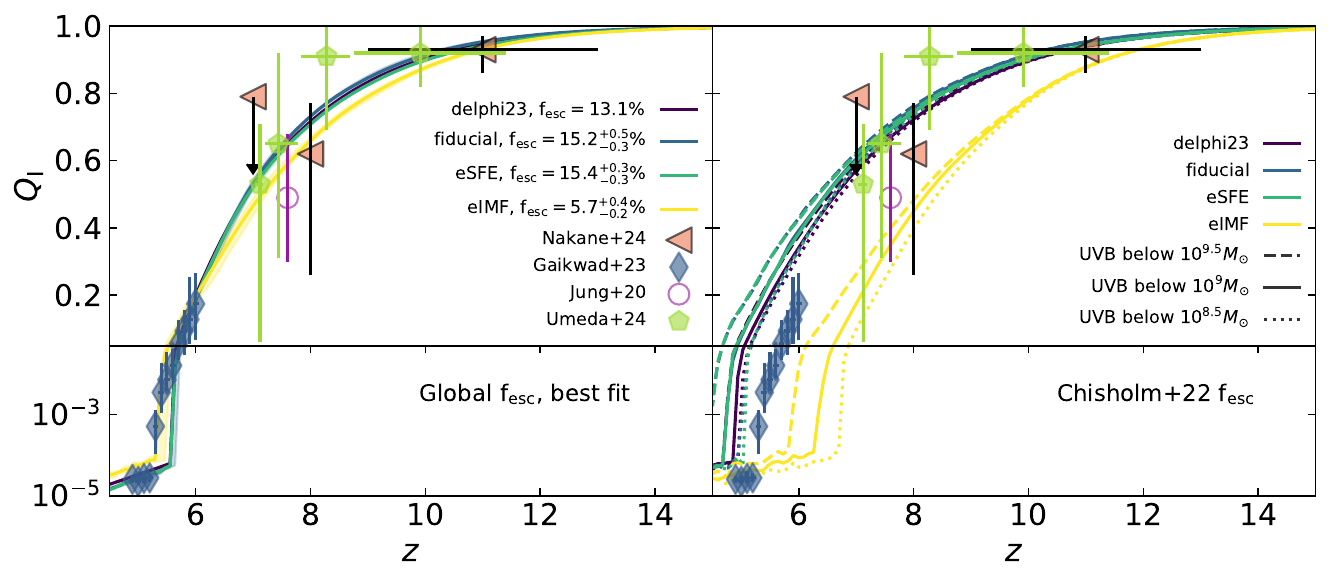}} 
\caption{Volume filling fraction of neutral hydrogen as a function of redshift. In the {\it left panel}, we carry out calculations assuming a constant global escape fraction ($\fesc$), as marked, given by the best fit to a selection of observational data \citep{Jung20, Gaikwad23, Nakane24, Umeda24}. Shaded areas represents the range of values obtained with the five runs using different random seeds. In the {\it right panel}, we show results using the escape fractions as in \cite{Chisholm22}, using our beta slopes as presented in Sect. \ref{sec:beta}. Solid, dashed, and dotted lines show results for the different halo masses ($10^{9}, 10^{9.5}$ and $10^{8.5}\Msun$) below which the baryonic content is completely suppressed due to the UVB. For clarity, shaded area are omitted in this panel, but would be similar as in the left one.}
  \label{fig:qhi_plot}
\end{center}
\end{figure*}

We then explore the cumulative contribution of galaxies of different stellar masses to the escaping emissivity of ionising photons in Fig. \ref{fig:cumul_nion}. For the sake of clarity, we omit the {\sc delphi}23 model and limit to showing results for a constant $\fesc$ value; we note that the results remain qualitatively unchanged when using the $\fesc$ relation from Eq. \ref{eq:chisholm_fesc}. 
At $z \sim 10$, in the fiducial and eSFE models, 50\% of ionising photons come from low-mass systems with $M_* \lsim 10^{6.75} \Msun$ and $M_* \lsim 10^7\Msun$ respectively; 90\% of escaping ionising photons can be traced to systems with $M_* \lsim 10^{8} ~(10^{8.5})\Msun$ in the fiducial and eSFE models, respectively. With its larger production rate of ionising photons per unit stellar mass, a larger fraction of the emissivity comes from lower-mass galaxies in the eIMF model: for example, at $z \sim 10$, 50\% (90\%) of the total ionising photons come from systems with $M_* \lsim 10^{6.3}~(10^{7.8})\Msun$. By $z \sim 7$, the fiducial and eSFE model are almost indistinguishable with $\sim50\% (90\%)$ of the emissivity being provided by galaxies with $M_* \lsim 10^8~(10^{9.4}) \Msun$. In the eIMF model, by $z \sim 7$, the emissivity again shifts to slightly lower-mass systems: $\sim50\% (90\%)$ of the emissivity being provided by galaxies with $M_* \lsim 10^{7.6}~(10^{9.1}) \Msun$. To summarise, in any of the models discussed here, galaxies with $M_* \lsim 10^9 \Msun$ provide the bulk ($~\sim 75\%$) of the escaping ionising emissivity for reionisation in agreement with our previous works \citep[e.g.][]{Dayal24}.

We then discuss the progress of reionisation through the evolution of $Q_I$ in Fig. \ref{fig:qhi_plot} for both a constant $\fesc$ (left panel) and the $\fesc-\beta$ relation detailed in Eq. \ref{eq:chisholm_fesc} (right panel). 
In the constant $\fesc$ model, that has been tuned to reproduce a reionisation history matching observational constraints including on the tail end of the process \citep{Jung20, Gaikwad23,Nakane24,Umeda24}, reionisation starts at $z \sim 13$, is 50\% complete by $z \sim 7$ and finishes by $z \sim 5.5$. In the second model, that uses $\fesc^{\rm{chisholm}}$, while the fiducial, eSFE and {\sc delphi}23 models match constraints at $z \gsim 7$, they predict a later end of the reionisation process (by $\Delta z \sim 0.2$) compared to $z \lsim 6$ data \citep{Gaikwad23}; we explore different threshold masses for reionisation feedback ranging between $10^{8.5}-10^{9.5}\Msun$, which do not affect our results sensibly. Finally, with its much higher escaping emissivity, the eIMF model leads to a very fast progress of reionisation. In this case, reionisation is 50\% over at $z \sim 8.5$, compared to at $z \sim 7$ for the other models, and finishes between $z \sim 5.8-6.8$ depending on the exact halo mass below which sources are suppressed due to reionisation feedback. We conclude that the escape fraction prescription of \cite{Chisholm22}, based on UV $\beta$ slopes, is not suitable to produce realistic reionisation histories within our framework. A fixed global escape fraction for all halos better reproduce observations. It is possible that the relation between $\beta$ slopes and $\fesc$ evolve with redshift, which would make the low redshift fits of \cite{Chisholm22} invalid during the epoch of Reionisation. Another possibility is that the lack of nebular continuum and the simple dust geometry of our model does not produce fully realistic $\beta$ slopes.

\section{Summary and discussion} \label{sec:summary}

We use the {\sc delphi} semi-analytic framework to study the galaxy populations being detected by the JWST in the first billion years of the Universe. Crucially, we 
extract ISM properties - namely the cold gas fraction and star formation efficiency - from {\sc sphinx}$^{20}$, a state-of-the-art radiation hydrodynamics simulation \citep{Rosdahl22}. Both the cold gas fraction ($\fcold$) and star formation efficiency ($f_*^{\rm{sphinx}}$) values are mass- and redshift-dependent. For example, at $z \sim 9$ while a bulk ($\sim 80\%$) of low-mass halos ($\mh \lsim 10^{8.4}\Msun$) show suppressed values of both ($\fcold \lsim 0.18$; $f_*^{\rm{sphinx}} \lsim 0.05$), both quantities increase with halo mass such that half of the most massive systems (with $\mh \gsim 10^{10}\Msun$) show $\fcold \sim 0.2-0.4$ and $f_*^{\rm{sphinx}} \sim 0.05-0.4$. This is a key first step in implementing physically motivated and tested prescriptions into semi-analytic models. In addition to inducing stochasticity in star formation, the fraction of gas in cold and warm phases is crucial for the dust calculations (ISM growth and dust destruction) included in our model that are base-lined against the latest ALMA results. 

While the augmentation with stochastic star formation allows us to reproduce the redshift evolution of the UV LF up to $z \sim 10$, this fiducial model under-predicts the bright end of the $z>10$ redshift Universe as probed recently by JWST. We explore two different approaches to match this observed overabundance of bright sources in the first billion years: in the eIMF model, the gas-phase metallicity- and redshift-dependent IMF becomes increasingly top-heavy at higher redshifts and lower metallicities as in previous works \citep[e.g.][]{Cueto24}. In the eSFE model, on the other hand, both the cold gas fractions and the star formation efficiencies of massive halos increase with increasing redshift; we maintain a Kroupa IMF as in the fiducial model. Both these models reproduce the 'dust attenuated' (observed) UV LF from $z \sim 5$ out to $z \sim 15-20$. Besides the UV LF, our models predict a wide range of observables at $z \sim 5-20$ including the stellar mass function, the dust mass-stellar mass relations, mass-metallicity relations, the UV spectral slopes ($\beta$), the Balmer break strength and the ionising photon production efficiency.
Our key findings are:

\begin{itemize}

\item[$\blacksquare$] Dust attenuation plays an important role in determining the bright end ($\MUV \lsim -21$) of the UV LF at $z \lsim 11$ in all of the models studied in this work. The impact of dust attenuation decreases with increasing redshift - this is due to dust and gas being more dispersed (covering a larger fractional volume of the host halo) with increasing redshift, leading to a decrease in the dust optical depth. Dust plays no role in attenuating the UV luminosity at $z \gsim 11-12$. 

\item[$\blacksquare$] The varying mass-to-light ratios in the different models explored in this work yield stellar mass functions that differ significantly, specially at the massive end at high redshifts; within observational dispersions, all models are in reasonable agreement with available data. For example, the number density associated with stellar mass of $M_* \sim 10^9\Msun$ at $z \sim 12$ varies by more than 2.5 orders of magnitude between the upper limit set by the eSFE model and the lower limit set by the eIMF model. A conjunction of the UV LF and SMF will be crucial in demarcating models predicting the correct mass-to-light ratios.    

\item[$\blacksquare$] The mass-metallicity relation is in place as early as $z \sim 17$ in all models. The fiducial model shows a metallicity value of about $2\% ~(36\%)\Zsun$ for $M_* \sim 10^{7}~ (10^{9.5-11.5})\Msun$ at $z \sim 5$. The eIMF model shows an increase in the mean amplitude of the mass-metallicity relation by about $0.2~(0.4)$ dex at $z \sim 9~(12)$ compared to the fiducial model, although the slopes remain similar. At $z \gsim 12$, the eSFE model shows the largest metallicity values. 

\item[$\blacksquare$] We find values of $\beta \lsim -2.3$ for systems fainter than UV magnitudes of about $-20$. Massive systems at $z\sim5$ reach $\beta$ slopes as red as $\sim-0.75$, but become increasingly bluer with increasing redshift.  

\item[$\blacksquare$] Galaxies in the eIMF model show the steepest Balmer break strength-$\MUV$ relation at $z \sim 6$, increasing from from $\sim 0.9$ to $1.6$ as $\MUV$ increases from -16 to -22. All the other models find relatively flat slopes with Balmer break strengths of about 1.4 at $z \sim 5$. 

\item[$\blacksquare$] Galaxies in the eIMF model show elevated values of $\xion$: for example, at $z \sim 5$ galaxies with $\MUV \gsim -17$ show values as high as $\log(\xion)=25.55~{\rm [Hz~erg^{-1}]}$ compared to values of about 25.2 in all other models; systems also saturate to $\log(\xion)=25.6~{\rm [Hz~erg^{-1}]}$ at $z \gsim 12$, a factor 2-2.5 times higher than those in all other models. These spectral signatures will be crucial in differentiating between models as more data comes in.

\item[$\blacksquare$] In all scenarios compatible with reionisation observables, star formation in low-mass galaxies ($M_* \lsim 10^9\Msun$) provide about 85\% of the 'escaping' emissivity for reionisation down to the midpoint at $z \sim 7$, irrespective of the model considered; higher mass systems contribute at most 30\% by the end stages of reionisation. Indeed, at $z \sim 10$, 90\% of the cumulative escaping ionising photons can be traced to systems with $M_* \lsim 10^{8-8.5}\Msun$ in the different models considered here. The (constant) values of $\fesc$ range between $5.7\%$ for the eIMF model to $\sim 15\%$ for the eSFE and fiducial models.

\end{itemize}

Given their dispersions, existing datasets are, as of now, equally compatible with the fiducial, eIMF and eSFE models. Further constraints on mass-to-light ratios (through joint analyses of the UV LF and SMF), spectral signatures (such as Balmer break strengths and the ionising photon production efficiency) and stronger hints on $\fesc$ \citep[e.g.][]{Choustikov24, Jaskot24, Jung24, Kerutt24} will be crucial in differentiating between these classes of models. 

Finally, our models can be further improved in the future to better reflect on the complex and diverse nature of high redshift galaxies. We have seen, for example, that our models do not recover extremely metal rich, low-mass systems, $\beta$ slopes bluer than -2.5 or Balmer break strengths weaker than 0.8. These could be driven by, for example, stochasticity in the star formation process where extremely young stars ($<5$ Myr) dominate the spectrum or the presence of very massive stars (up to 400 $\Msun$) boosting the luminosity of star-forming galaxies \citep{schaerer2025}. Our metal prescriptions use the instantaneous recycling approximation that could be responsible for the lack of low-mass high-metallicity systems \citep[Sect. 4.1;][]{Ucci2023}. Further we assume perfect mixing of gas and dust and assume a simple dust geometry to calculate the total optical depth, ignoring both clumping of dust and the spatial segregation between star forming regions and dust \citep{Inami22}. We aim at including these crucial effects into our models in future works to keep pace with the incredible datasets available in the first billion years.




\begin{acknowledgements}

VM and PD acknowledge support from the NWO grant 016.VIDI.189.162 ('ODIN'). PD warmly thanks the European Commission's and University of Groningen's CO-FUND Rosalind Franklin program. MGH has been supported by STFC consolidated grants ST/N000927/1 and ST/S000623/1. TK was supported by the National Research Foundation of Korea (RS-2022-NR070872 and RS-2025-00516961) and by the Yonsei Fellowship, funded by Lee Youn Jae. The authors thank S. Gazagnes, A. Hutter, J. Kerutt, C. Leiterer, and M. Trebitsch for their helpful comments and insightful discussions.

\end{acknowledgements}

\bibliography{references}
\bibliographystyle{aa} 

\end{document}